\documentclass[preprint,nofootinbib]{revtex4-1}
\usepackage{graphicx,epsfig,rotate,hyperref}

\def\iso#1#2{\mbox{${}^{#2}{\rm #1}$}}
\def\he#1{\iso{He}{#1}}
\def\li#1{\iso{Li}{#1}}
\def\be#1{\iso{Be}{#1}}
\def\dpg{$d(p,\gamma)$\he3~}

\def\e10{\eta_{10}}

\def\Ombh2{\Omega_{\rm b} h^2}
\def\omb{\omega_{\rm b}}

\def\ga{\mathrel{\raise.3ex\hbox{$>$\kern-.75em\lower1ex\hbox{$\sim$}}}}
\def\la{\mathrel{\raise.3ex\hbox{$<$\kern-.75em\lower1ex\hbox{$\sim$}}}}

\def\beq{\begin{equation}}
\def\eeq{\end{equation}}
\def\beqar{\begin{eqnarray}}
\def\eeqar{\end{eqnarray}}

\begin{document}

\rightline{UMN--TH--4004/20}
\rightline{FTPI--MINN--20/35}
\rightline{November 2020}

\title{The Impact of New $d(p,\gamma)\he3$ Rates on Big Bang Nucleosynthesis}

\author{Tsung-Han Yeh}
\affiliation{Department of Physics, University of Illinois,~
Urbana, IL 61801}
\author{Keith A. Olive}
\affiliation{William I. Fine Theoretical Physics Institute,
School of Physics and Astronomy,
University of Minnesota, Minneapolis, MN 55455, USA}
\author{Brian D. Fields}
\affiliation{Departments of Astronomy and of Physics, University of Illinois,
Urbana, IL 61801}

\begin{abstract}
  We consider the effect  on Big Bang Nucleosynthesis (BBN) of new measurements of the 
  $d(p,\gamma)\he3$ cross section by the LUNA Collaboration. These have an important effect on the primordial abundance of D/H which is also sensitive to the baryon density at the time of BBN. 
  We have re-evaluated the thermal rate for this reaction, using
  a world average of cross section data, which we describe with
  model-independent polynomials; our results are in good agreement with a similar
  analysis by LUNA.
  We then perform a full likelihood analysis
  combining BBN and {\em Planck} cosmic microwave background (CMB) likelihood chains 
  using the new rate combined with previous measurements and compare with the results using previous rates. Concordance between BBN and CMB measurements of the anisotropy spectrum using the old rates was excellent. The predicted deuterium abundance at the Planck value of the baryon density was $({\rm D/H})_{\rm BBN+CMB}^{\rm old} = (2.57 \pm 0.13) \times 10^{-5}$ which can be compared with the value determined from quasar absorption systems $({\rm D/H})_{\rm obs} = (2.55 \pm 0.03) \times 10^{-5} $. Using the new rates we find $({\rm D/H})_{\rm BBN+CMB} = (2.51 \pm 0.11) \times 10^{-5} $.
  We thus find consistency among BBN theory, deuterium and \he4 observations, and the CMB, when using reaction rates fit in our data-driven approach.  We also find that the new reaction data tightens the constraints on the number of relativistic degrees of freedom during BBN, giving the effective number of light neutrino species $N_\nu = 2.880 \pm 0.144$
  in good agreement with the Standard Model of particle physics.
  Finally, we note that the observed deuterium abundance continues to be more precise than the BBN+CMB prediction, whose error budget is now
  dominated by $d(d,n)\he3$ and $d(d,p){}^{3}{\rm H}$.
  More broadly, it is clear that the details of
  the treatment of nuclear reactions and their uncertainty have become critical for BBN.
\end{abstract}


\maketitle

\section{Introduction}

Big Bang Nucleosynthesis (BBN) is one of the deepest probes in the early universe which uses known physics \cite{bbn,cfo1,bn,coc,cfo2,cfo3,coc03,cuoco,cyburt,coc04,iocco,pisanti,coc3,cfo5,coc4,coc15,coc18,CFOY,foyy}. Standard BBN is formulated in the context of a $\Lambda$CDM cosmology, standard nuclear and particle physics with three light neutrino flavors. Given the reasonably accurate value of the neutron mean-life, $\tau_n = 879.4 \pm 0.6$ s \cite{rpp}, and the determination of the baryon density, $\omega_B = \Omega_B h^2$, (or baryon-to-photon ratio, $\eta = n_{\rm B}/n_\gamma$) from the cosmic microwave background (CMB) anisotropy spectrum \cite{Planck2018}, BBN is effectively a parameter-free theory \cite{cfo2}.  The theory predicts the primordial abundances of the light element isotopes, D, \he3, \he4, and \li7. While there is no
direct comparison to an observation of primordial \he3, 
the other three isotopes can be tested by observations.
Despite the well known discrepancy\footnote{It is quite unlikely \cite{cfo4,boyd,brog,ic20} that the solution to the lithium problem resides in the underlying nuclear physics. The discrepancy may be a result of stellar depletion \cite{dep} and there is some evidence of a broken lithium plateau at very low metallicity \cite{broken}.
We do not consider the lithium problem further here.}  with \li7 \cite{cfo5,fieldsliprob}, both \he4 and D are in excellent agreement with observations. 

Rather than simply computing the light elements abundances assuming a particular value of $\eta$ quoted by Planck, we have, in previous work \cite{CFOY,foyy}, advocated a likelihood approach which convolves the Planck likelihood chains, with a Monte-Carlo simulation of nuclear cross sections. In \cite{foyy}, hereafter FOYY, the combined BBN+CMB likelihood functions for \he4 and D/H
are characterized by a mean and width given by
$Y_p = 0.2469 \pm 0.0002$ and D/H $\times 10^5 = 2.57 \pm 0.13$
where $Y_p$ is the primordial \he4 mass fraction, and D/H is the abundance of deuterium by number relative to hydrogen. These should be compared with the observational determinations, $Y_p = 0.2453 \pm 0.0034$ \cite{abopss}
\footnote{This value is an update of the value used in FOYY, $Y_p = 0.2449 \pm 0.0040$ \cite{aos4}.} and
 \beq
 \label{eq:Dobs}
({\rm D/H})_{\rm obs} = (2.55 \pm 0.03) \times 10^{-5}
 \eeq
 based on an weighted average of 11 measurements \cite{pc,cooke,riemer,bala,cookeN,riemer17,zava,CPS}.  As one can see, both observations are in excellent agreement with BBN predictions.

The BBN predictions of the light element abundances rely on a detailed network of nuclear reaction rates, thermally averaged over the energy range of importance to BBN. Each of the these rates carries an uncertainty,
which is used in a Monte-Carlo simulation to produce
likelihood distributions of each of the light element abundances. Clearly accurate measurement of nuclear cross sections are critical for obtaining accurate light element abundances. 
 Indeed, the deuterium observational error budget is far tighter than the 5\% level
 found when combining BBN theory (pre-LUNA) and the CMB \cite{foyy}.

Among the processes of particular importance for calculating the D/H abundance is $d(p,\gamma)$\he3.
Our nominal rate for this process is based on an average of several experimental measurements compiled by NACRE \cite{nacreII} in the energy range of interest to BBN.
Nollett and collaborators have emphasized \cite{Nollett2011,bbnt,cookeN} that theoretical calculations predict a cross section 20\% larger than the measurements in the BBN energy range. These calculations are derived from an {\em ab initio} quantum mechanical calculation using nucleon interaction potentials, and agree with $d(p,\gamma)\he3$ data at energies outside of the BBN window.  
An independent calculation by Marcucci and collaborators \cite{marcucci2005}, updated in \cite{marc}, finds a similar 
mismatch at BBN energies.
Another theory-based approach is the $R$-matrix analysis
was used by the Paris group \cite{coc15}.  This also gives a higher BBN cross section and rate. The higher rate leads to a lower abundance of D/H, and a primordial value of D/H $\times 10^5 = 2.45 \pm 0.05$ was found \cite{coc15}, updated to D/H $\times 10^5 = 2.46 \pm 0.04$ \cite{coc18}, both in evident disagreement with observations. 

Thus it has become clear that new, precision $d(p,\gamma)\he3$ data is needed, at energies targeted for BBN.
As a result, dedicated experiments have been conducted.   Ref.~\cite{tisma2019} first presented new results in the BBN energy
range, finding cross sections between the older data and the theoretical predictions.  These
used a thick-target method, with an error budget somewhat larger than the older data.
Meanwhile, an intensive study was initiated at the Laboratory for Underground Nuclear Astrophysics (LUNA)
in the Gran Sasso National Laboratory \cite{trezzi2018}.
Commissioning of the LUNA  $d(p,\gamma)\he3$ experiment was
 described in \cite{mossa1} and results have now been released \cite{mossa2}. 
 As we will see in more detail, the new LUNA data cover the entire BBN energy range with good sampling.
Just as importantly, they feature a tight systematic and statistical error budget.  
The measured LUNA cross section shows a slight increase compared to earlier measurements in the BBN energy range, 
but remains significantly below the theoretical calculations.
This long-anticipated result 
marks a major advance in understanding this rate and indeed all of BBN, and motivates our study here.
 
 In this note, we first include in section \ref{sec:rate} the newly measured $d(p,\gamma)$\he3
 to update the results of FOYY. In section \ref{sec:N3}, we compare the our previous results, with those replacing the $d(p,\gamma)$\he3 rate with the new experimental result \cite{mossa2} as
 well as our new baseline result based on averaging
 the new measurement with existing cross section measurements. We also compare these results with the calculated abundances using the theoretical rate for \dpg \cite{marc}. 
 We find that the world averaged cross-section is only slightly reduced relative to the new measurement, but the uncertainty is substantially reduced
due to the significant improvement of the measurement precision \cite{mossa2}. While we find 
that the BBN-CMB predicted value for D/H is decreased to $(2.51 \pm 0.11) \times 10^{-5}$, it remains consistent with observations
and remains significantly larger than the result based on the theoretical cross section which yields $(2.42 \pm 0.10) \times 10^{-5}$.
We also discuss in section \ref{sec:Nnot3} the results when the number of neutrino degrees of freedom, $N_\nu$, is not held fixed at 3.
We summarize the impact of our results in section \ref{sec:summary}. 

\section{The $d(p,\gamma){}^{3}{\rm He}$ Cross Section and Rate}
\label{sec:rate}

\begin{center}
  \begin{figure}[htb]
 \includegraphics[width=0.475\textwidth]{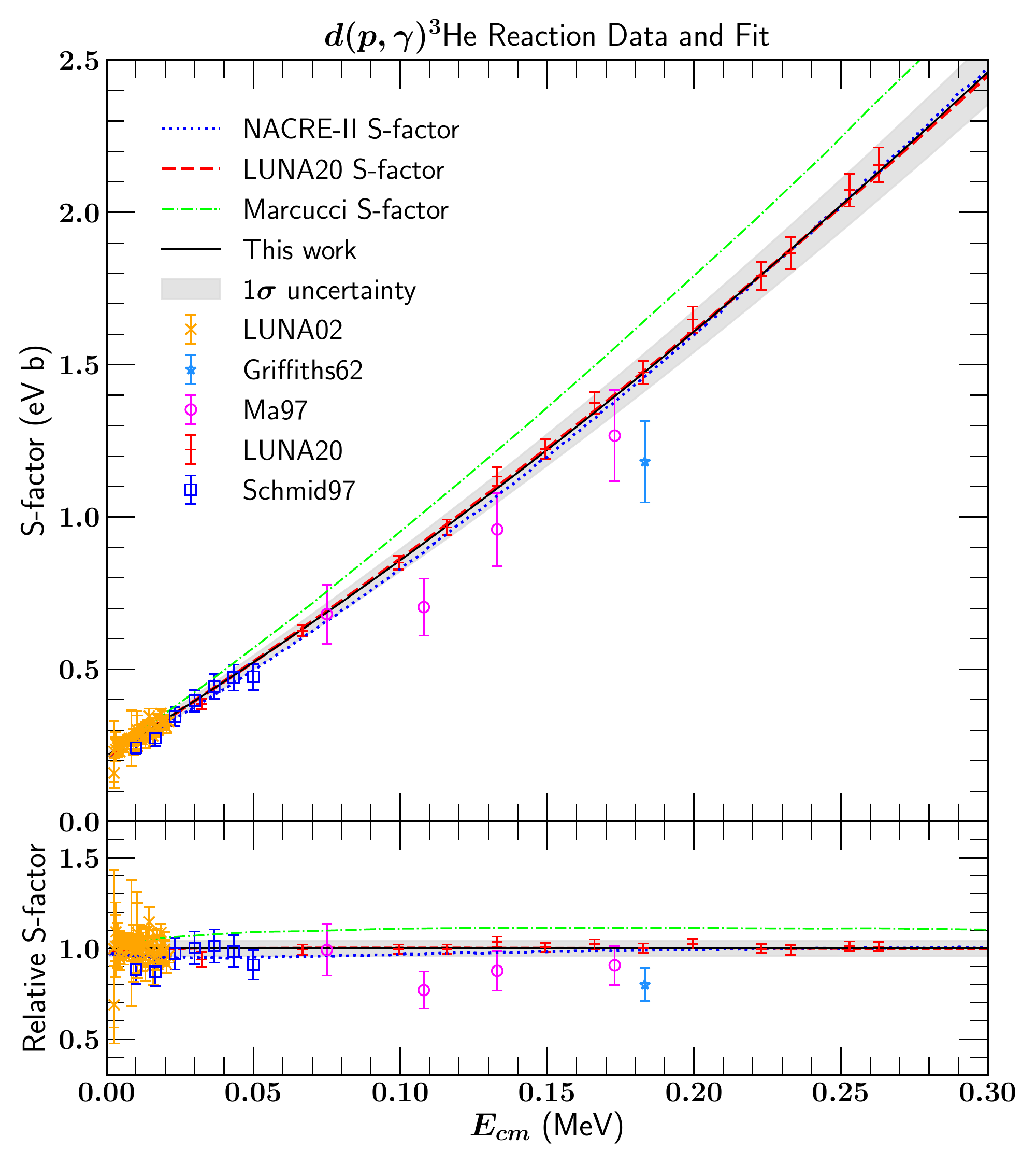}
  \includegraphics[width=0.475\textwidth]{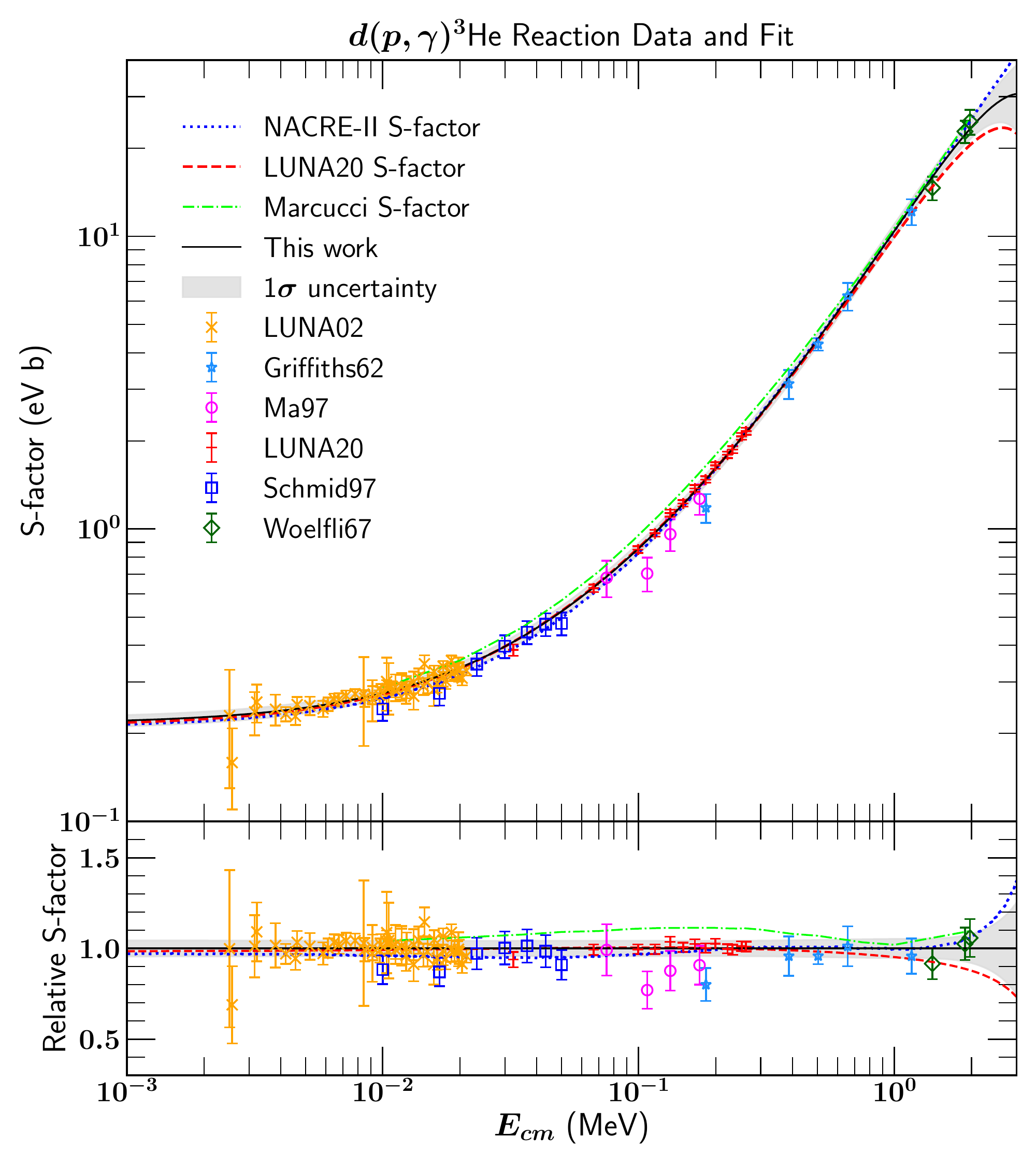}
    \caption{ The astrophysical S-factor for $d(p,\gamma)$\he3 showing 1) the NACRE-II \cite{nacreII} $S$-factor used in FOYY (blue dotted); 2) the theoretical $S$-factor \cite{marc} (green dot-dashed); 3) the LUNA global average \cite{mossa2} (red dashed); and 4) our new world average rate (black solid). The shading corresponds to the 68\% uncertainty we assign to the average rate. In the left panel, the S-factor is shown against a linear energy scale centered on the BBN energies. In the right panel, we show an extended energy range on a log scale. }
    \label{fig:sfac}
  \end{figure}
\end{center}

We have performed a new calculation of the \dpg\ thermonuclear rate and its uncertainty.  To do this, we first evaluate the \dpg\ cross section.  
We use cross section data from LUNA \cite{mossa1,mossa2},
as well as earlier work from 1997 by Schmid et al.~\cite{Schmid1997} and Ma et al.~\cite{Ma1997}, and from the 1960s by W\"olfli et al.~\cite{Wolfli1967} and Griffiths et al.~\cite{Griffiths1962}.\footnote{
We do not use the results of refs.~\cite{Bailey1970} nor~\cite{Griffiths1963} due to systematic issues
with the stopping powers they used, as pointed
out by both Ma \cite{Ma1997} and Schmid \cite{Schmid1997}.}
These data are plotted in Fig.~\ref{fig:sfac} 
in terms of the astrophysical
$S(E) = E \ \sigma(E) \ e^{2\pi \eta_{\rm s}}$
where for this reaction the usual barrier penetration exponent is
$2\pi \eta_{\rm s} = 2\pi e^2/\hbar v$, with $v(E)$
the relative velocity.
We see that the $S$-factor is a smooth function of energy,
and nearly linear in the BBN energy range.

Following a procedure similar to Cyburt \cite{cyburt} and FOYY,
we perform a global fit for $S(E)$ over the energy range $2 \ \rm keV$
to $2 \ \rm MeV$, which encompasses the BBN range.
We describe $S(E)$ with a polynomial in $E$, and we find
that a 3rd-order expansion is sufficient
to characterize the rate without overfitting.
We combine the statistical and
systematic errors in quadrature to obtain and energy-dependent
error, and use it to find a global best fit and associated uncertainty band arising from minimizing $\chi^2$.  This amounts to
a weighted average favoring measurements like LUNA with
small statistical and systematic errors.  We include an energy-independent discrepancy error 
based that accounts for systematic differences between data sets;
this amounts to requiring an overall $\chi^2$ per degree of freedom
no larger than unity.

For comparison, in FOYY we used NACRE-II \cite{nacreII} for this reaction;
their global fit of course did not include LUNA, but
also used a larger data set that extended to higher energies
far outside of the BBN range.
LUNA performed their own global fit, using a 
polynomial-based procedure very similar to ours.
Finally, the theory-based {\em ab initio}
predictions of \cite{marc}
do not directly use \dpg\ but do include an error analysis
based on the uncertainties in the model inputs;
in using their result we combine their two error terms in quadrature.

Figure~\ref{fig:sfac} shows
the astrophysical $S$-factor for $d(p,\gamma)$\he3 
with the measurements and fits
considered in this work,
plotted as a function of energy displayed on a linear scale (left) and an expanded log scale (right).  The curves show
the fits we considered, corresponding to 
 1) the NACRE-II \cite{nacreII} $S$-factor used in FOYY (blue dotted); 2) the theoretical $S$-factor \cite{marc} (green dot-dashed); 3) the LUNA global average \cite{mossa2} (red dashed); and 4) our newly world average rate (black solid).  
 As one can see, our world averaged result (which we take to be our current baseline) is very similar to the new LUNA measurement
 and lies between the NACRE-II average and the theoretical calculation (though the new result is much closer to the former).

For each set of $S(E)$ factors shown in Fig.~\ref{fig:sfac}, we compute a thermal average in the usual manner to obtain a thermonuclear rate coefficient as a function of temperature:
\beq
\lambda(T) = N_{\rm Avo} \langle \sigma v \rangle
=  m_{\rm u}^{-1} \,  \left( \frac{8}{\mu \pi} \right)^{1/2} \, (kT)^{-3/2} \, 
 \int_{E_{\rm min}}^{E_{\rm max}} S(E) \  e^{-2\pi \eta} \ e^{-E/kT}
\ dE
\eeq
where Avogadro's number $N_{\rm Avo} = 1/m_{\rm u}$ is also
the inverse of the atomic mass unit $m_{\rm u}$, and $\mu$ is the reduced mass.
We choose integration limits $(E_{\rm min},E_{\rm max}) = (0, 100T)$.

\begin{center}
  \begin{figure}[htb]
 \includegraphics[width=0.95\textwidth]{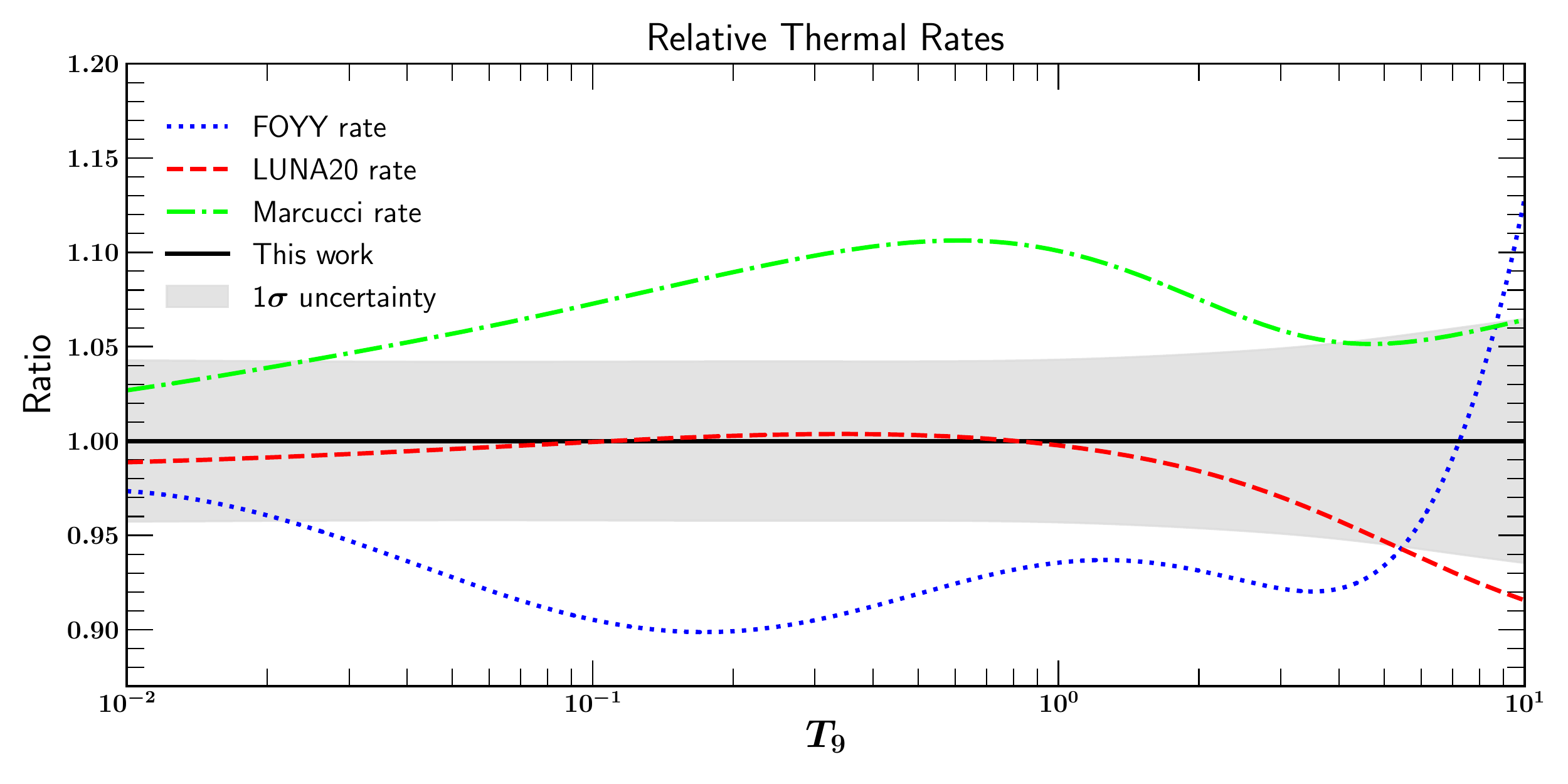}
    \caption{The relative difference between the rates compared to our new baseline average.
    Shown here as a function of temperature in units of $10^9$ K.}
    \label{fig:diff}
  \end{figure}
\end{center}

The relative difference between the thermonuclear rates is shown in Fig.~\ref{fig:diff}. We set our baseline average (corresponding to the black solid curve in Fig.~\ref{fig:sfac}) to 1 and display the other three rates relative to this. We see that compared to the rate used in FOYY based on the NACRE-II $S$-factor, the rate has increased at all temperatures in the BBN range, with about a 6.5\% increase around $T=10^9 \ \rm K$ where the rate is most important.  On the other hand, our rate remains $\sim 10\%$ lower than the theoretical prediction.  Finally, our results are quite similar to the LUNA global average, reflecting our similar approaches, and indeed they are almost identical around and below the $T = 10^9 \ \rm K$ regime important for BBN.  

\section{Results for fixed $N_\nu = 3$}
\label{sec:N3}
 
 To describe the impact of the newly measured $d(p,\gamma)$\he3 cross section, we first briefly review our method for constructing BBN likelihood functions.
 As uncertainties in the input BBN reactions
play a pivotal role in determining
the uncertainties in the light element abundances, we apply Monte-Carlo techniques \cite{kr,kk,hata,fo} for constructing likelihood functions. These have been discussed in detail in \cite{CFOY,foyy}, so here we summarize briefly. 

We generate likelihoods for BBN light element predictions via a Monte Carlo approach.
For the case of $N_\nu=3$ we run our BBN code for a grid in $\eta$;
when $N_\nu$ is allowed to vary we run for a grid in $(\eta,N_\nu)$.
For each grid point, i.e., each value of $\eta$ and $N_\nu$, we run the BBN code 10000 times
allowing the key nuclear rates to vary according to their uncertainties.
Each rate is drawn from a Gaussian distribution with the appropriate mean and uncertainty.
From the resulting light element abundances at each $(\eta,N_\nu)$ grid point we compute the mean
and uncertainties. From these we infer BBN likelihood functions for the entire grid.
In the case of fixed $N_\nu=3$, this gives ${\mathcal L}_{\rm BBN}(\eta;X)$
with abundances $X \in (Y_p,{\rm D/H})$, and for $N_\nu$ varying we compute
${\mathcal L}_{\rm NBBN}(\eta,N_\nu;X)$.

The CMB independently measures $\eta$ to an exquisite precision, and now also provides important constraints on $Y_p$ and $N_\nu$.
As noted above, we define a CMB likelihood function,
derived from 
the Markov chains generated from the final {\it Planck} 2018
data release \cite{Planck2018}.\footnote{We use the {\tt base\_yhe} and {\tt base\_nnu\_yhe} chains in \href{https://wiki.cosmos.esa.int/planck-legacy-archive/index.php/Cosmological_Parameters}{https://wiki.cosmos.esa.int/planck-legacy-archive/index.php/Cosmological\_Parameters}}
As in FOYY, we use the CMB
chains based on temperature and polarization data, TT+TE+EE+lowE,
including CMB lensing. The chains we employ do not assume any BBN relation between the  baryon density and helium abundance. We separately consider the cases where $N_\nu = 3$ is fixed, and where it is free to vary; 
the resulting likelihoods are called ${\mathcal L}_{\rm CMB}(\eta,Y_p)$ and ${\mathcal L}_{\rm NCMB}(\eta,Y_p,N_\nu)$ respectively.
These can be convolved with
the BBN likelihood functions defined above.  
In addition, we have the observational likelihoods ${\mathcal L}_{\rm Obs}(X)$ which are assumed to be Gaussians based on the observational determinations 
for $X \in (Y_p,{\rm D/H})$.

We begin with the construction of the BBN-CMB likelihood functions for the light element abundances. For now, we
will assume $N_\nu = 3$ and consider the more general case below. The combined likelihood is defined by
\beq
{\mathcal L}_{\rm BBN+CMB}(X_i) \propto \int 
  {\mathcal L}_{\rm CMB}(\eta,Y_p) \
  {\mathcal L}_{\rm BBN}(\eta;X_i) \ d\eta \, ,
\label{CMB+BBN}
\eeq
where we normalize each of the likelihood functions so that at their peaks $\mathcal{L}=1$.

\begin{center}
\begin{figure}[!htb]
\includegraphics[width=0.80\textwidth]{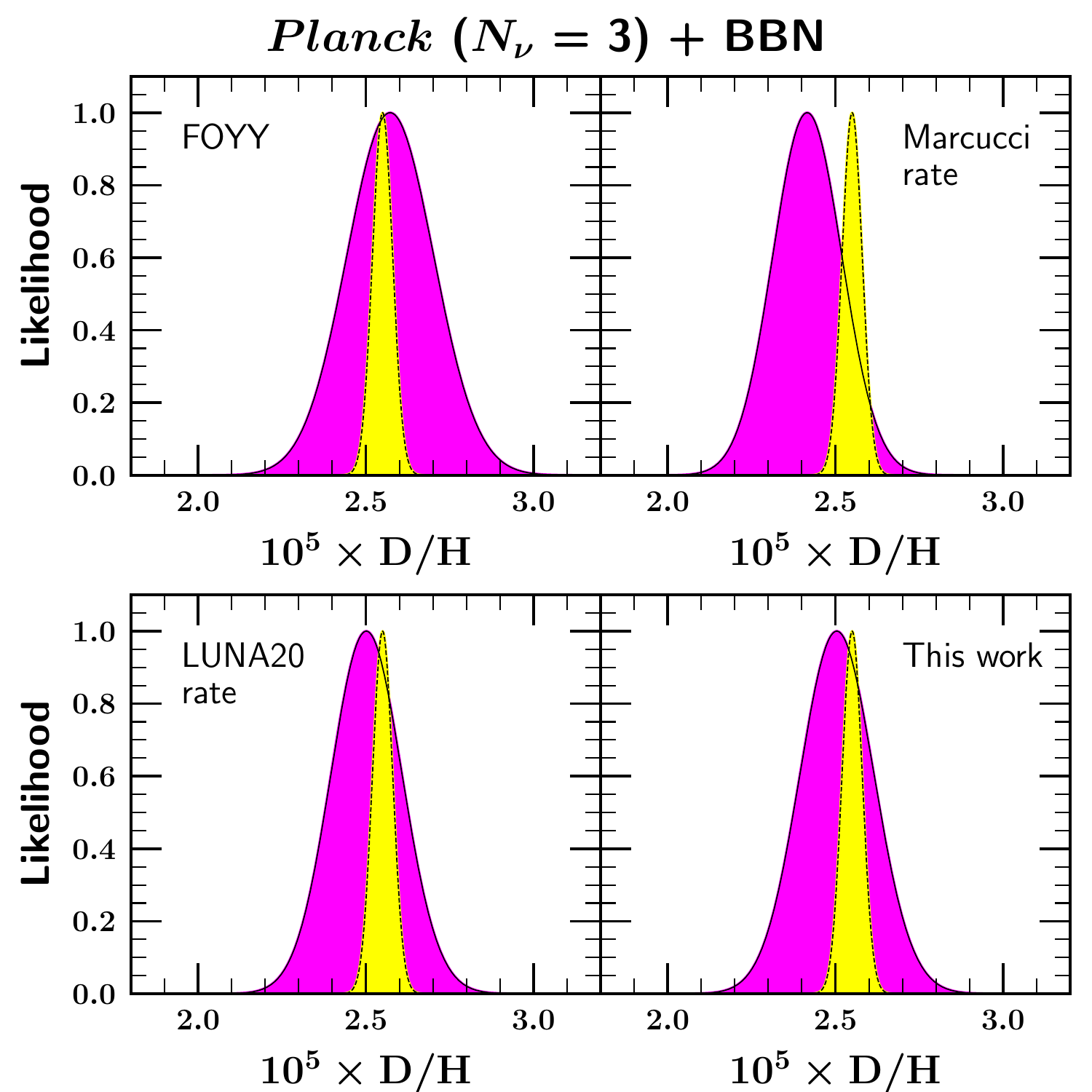}
\caption{
Likelihood functions for D/H
with different assumed rates for \dpg: 
i) the  FOYY rate (upper left), ii) the theoretical rate suggested in \cite{marc} (upper right), iii) the newly measured rate from \cite{mossa2} (lower left), and iv) our new combined rate
including the results of \cite{mossa2} (lower right).
The solid-lined,
dark-shaded (purple) curves are 
the BBN+CMB predictions, based on {\em Planck} inputs as discussed
in the text.  The dashed-lined, light-shaded 
(yellow) curves show measurements from quasar absorption systems. 
\label{fig:2x2abs_2d}
}
\end{figure}
\end{center}

We show in Fig.~\ref{fig:2x2abs_2d}
the D/H likelihood as defined in Eq.~(\ref{CMB+BBN})
using four choices of the rate for \dpg and shown in Fig.~\ref{fig:sfac} : i) the rate used in FOYY (upper left), ii) the theoretical rate suggested in \cite{marc} (upper right), iii) the newly measured rate from \cite{mossa2} (lower left), and iv) our new combined rate
including the results of \cite{mossa2} (lower right).
The combined BBN+CMB likelihood from Eq. (\ref{CMB+BBN}) is shaded purple. The observational likelihood is shaded yellow. 
As we have claimed previously \cite{CFOY,foyy}, when using the FOYY rate for \dpg, we get near perfect agreement between the BBN-CMB calculated deuterium abundance ($(2.574 \pm 0.129) \times 10^{-5}$) and the observational determination 
(eq.~\ref{eq:Dobs})
as seen in the upper left panel. In contrast, using the theoretical rate \cite{marc} gives
a mean abundance of $(2.417 \pm 0.103) \times 10^{-5}$ and leads to a shift to the left of the theoretical likelihood (purple). 
Using either the LUNA rate \cite{mossa2}, or our averaged rate gives a virtually identical result,
$(2.503 \pm 0.106) \times 10^{-5}$ and $(2.506 \pm 0.110) \times 10^{-5}$ as seen in the lower panels of Fig.~\ref{fig:2x2abs_2d}. These remain in excellent agreement with observations.

The BBN+CMB likelihoods in Fig.~\ref{fig:2x2abs_2d}
are summarized by
the predicted abundances for D/H given in Table~\ref{tab:DH},
where the central values give the mean,
and the error gives the $1\sigma$ variance.
The third column gives the value at the peak of the distribution. Each of these results can be compared to the observed abundance in eq.~(\ref{eq:Dobs}). 

\begin{table}[!htb]
\caption{The mean and peak values of D/H for each of the 
adopted rates for \dpg.
\label{tab:DH}
}
\begin{tabular}{|l|c|c|}
\hline
 \dpg rate & mean D/H $\times 10^5$ & peak D/H $\times 10^5$ \\
\hline
FOYY~\cite{foyy} & $2.574\pm0.129$ & $2.572 $ \\
\hline
Theory~\cite{marc} & $2.417\pm 0.103$ & $2.416$ \\
\hline
LUNA20~\cite{mossa2} & $2.503\pm 0.106$ & $2.502$ \\
\hline
\hline
This Work & $2.506\pm 0.110$ & $2.504$ \\
\hline
\end{tabular}
\end{table}

\begin{center}
  \begin{figure}[htb]
 \includegraphics[width=0.95\textwidth]{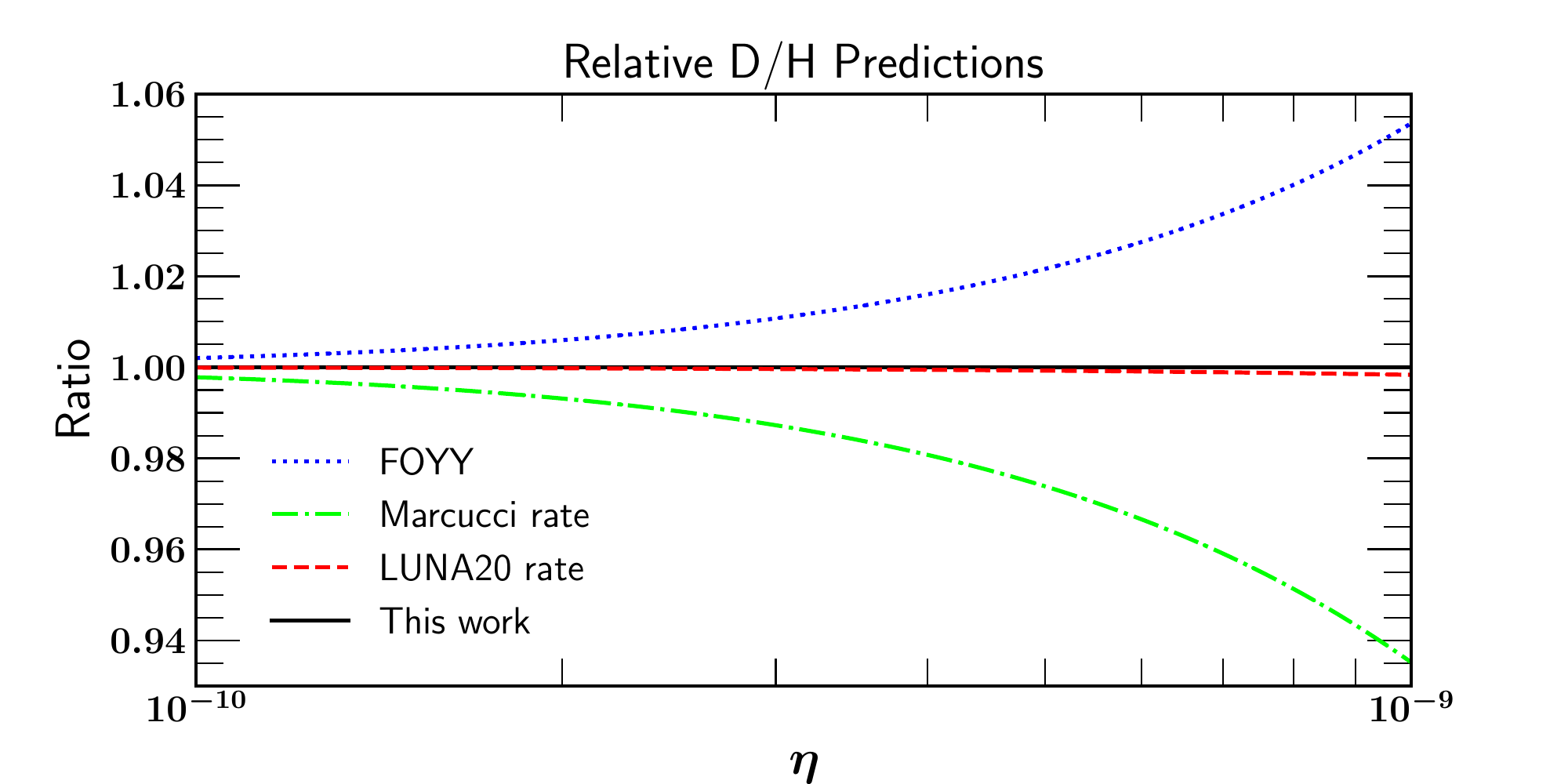}
    \caption{The relative change in the D/H abundance using our previous (FOYY) rate
    as a function of $\eta$. }
    \label{releta}
  \end{figure}
\end{center}

The relative change in D/H is shown in Fig.~\ref{releta} as a function of $\eta$. 
The D/H abundance using our current averaged
rate is set at 1
and the reductions or enhancement in D/H for the other three rates are shown as labelled. 
Once again, we see the very small difference in the resulting D/H between the rate
calculated here, and the one provided by LUNA \cite{mossa2}. 

\begin{center}
\begin{figure}[!htb]
\includegraphics[width=0.80\textwidth]{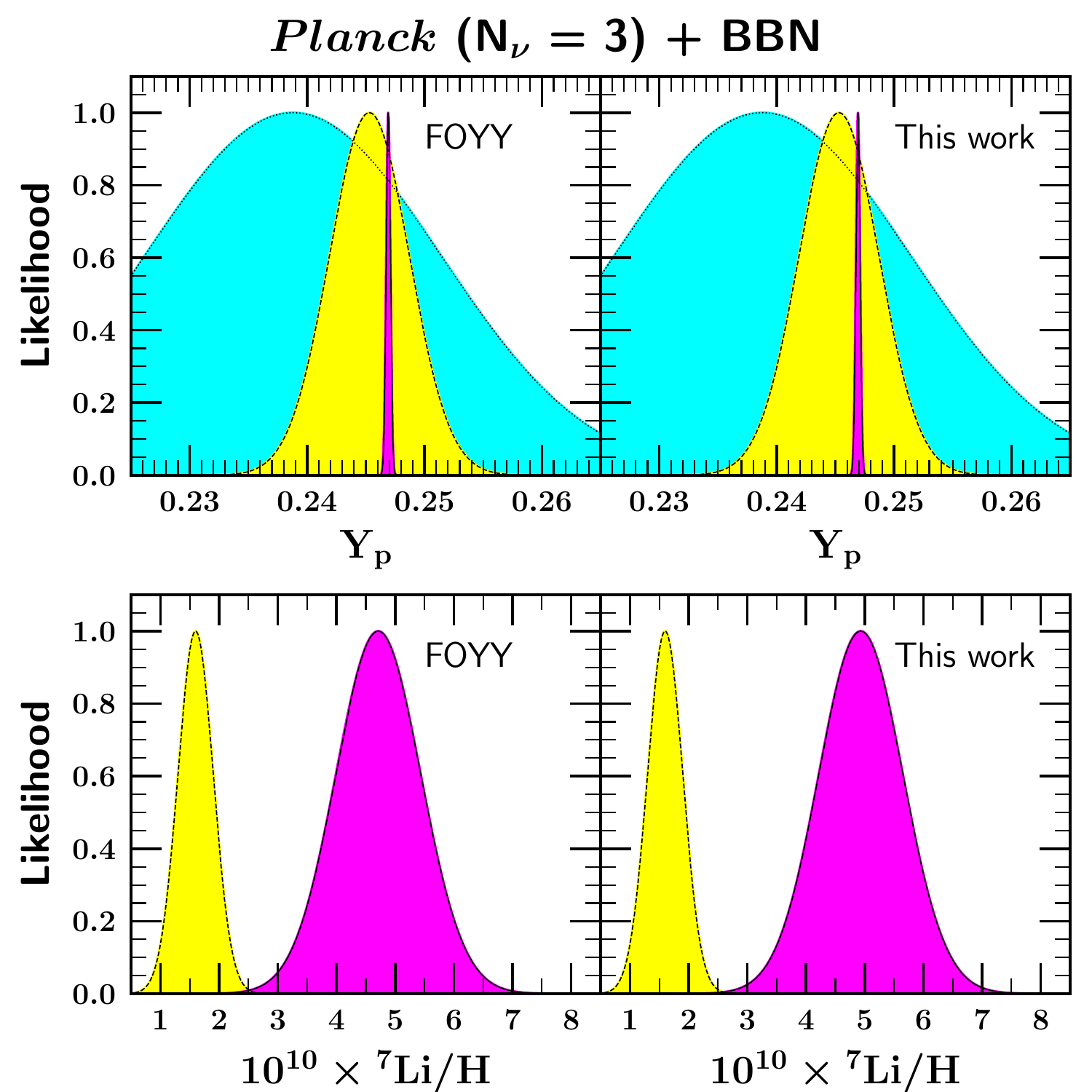}
\caption{
Likelihood functions for \he4 (upper panels) and \li7 (lower panels). Here we consider only the FOYY \dpg rate (left panels) and 
our new combined rate
including the results of \cite{mossa2} (right panels)
The solid-lined,
dark-shaded (purple) curves are 
the BBN+CMB predictions, based on {\em Planck} inputs as discussed
in the text.  The dashed-lined, light-shaded 
(yellow) curves show the observational likelihood functions. For \he4, the dotted-lined, medium- shaded (cyan) curve shows the independent CMB determination of \he4.
\label{fig:2x2abs_heli}
}
\end{figure}
\end{center}

As the \dpg rate also affects \he4 and \li7,
we show in Fig.~\ref{fig:2x2abs_heli}
the likelihood functions for \he4 (upper panels) and \li7 (lower panels), comparing the previous results from FOYY (left panels) with the combined results which include the newly measured 
\dpg rate (right panels). 
In the case of \he4, we also show the likelihood 
obtained by integrating ${\mathcal L}_{\rm CMB}(\eta,Y_p)$ over $\eta$ corresponding to the Planck CMB determination of $Y_p$. As one might expect, there is virtually no change in the helium abundance. In contrast there is a more noticeable change in Li/H towards higher values, thus worsening the lithium problem (slightly). 

The predicted abundances for \he4 and \li7 from Fig.~\ref{fig:2x2abs_heli}
in addition to the remaining two rates
are summarized in Table~\ref{tab:YLi}.

\begin{table}[!htb]
\caption{The mean and peak values of $Y_p$ and Li/H for each of the 
adopted rates for \dpg.
\label{tab:YLi}
}
\begin{tabular}{|l|c|c|}
\hline
 \dpg rate & mean $Y_p$ & peak $Y_p$ \\
\hline
FOYY~\cite{foyy} & $0.24691\pm0.00018$ & $0.24691 $ \\
\hline
Theory~\cite{marc} & $0.24693\pm0.00018$ & $0.24693$ \\
\hline
LUNA20~\cite{mossa2} & $0.24693\pm0.00018$ & $0.24693$ \\
\hline
\hline
This Work & $0.24693\pm0.00018$ & $0.24693$ \\
\hline
\hline
 \dpg rate & mean Li/H $\times 10^{10}$ & peak Li/H $\times 10^{10}$ \\
\hline
FOYY~\cite{foyy} & $4.72\pm 0.72$ & $4.71$ \\
\hline
Theory~\cite{marc} & $5.26 \pm 0.75$ & $5.26$ \\
\hline
LUNA20~\cite{mossa2} & $4.94 \pm 0.71$ & $4.94$ \\
\hline
\hline
This Work & $4.94\pm 0.72$ & $4.93$ \\
\hline
\end{tabular}
\end{table}

We can marginalize 
any of our likelihood functions to obtain a likelihood as a function of $\eta$ alone, $\mathcal{L}(\eta)$. We note that the CMB-alone value for $\eta_{10} = \eta \times 10^{10}$ is relatively low, $6.104 \pm 0.058$ \cite{foyy}. When the BBN relation between $Y_p$ and $\eta$ is included, the best fit value for $\eta_{10}$ is higher, $6.129 \pm 0.040$. 
Although D/H is commonly
referred to as an excellent baryometer, the still relatively large uncertainty in the calculated D/H abundance does not affect the likelihood in a significant way. From the upper left panel of Fig.~\ref{fig:2x2abs_2d}, we see that the observed D/H abundance is lower than the calculated value and should draw the best fit value of
$\eta_{10}$ upwards. It does, but only marginally so, and within round-off it remains at $6.129 \pm 0.040$. 

\begin{figure}[!htb]
    \centering
    \includegraphics[width=0.51\textwidth]{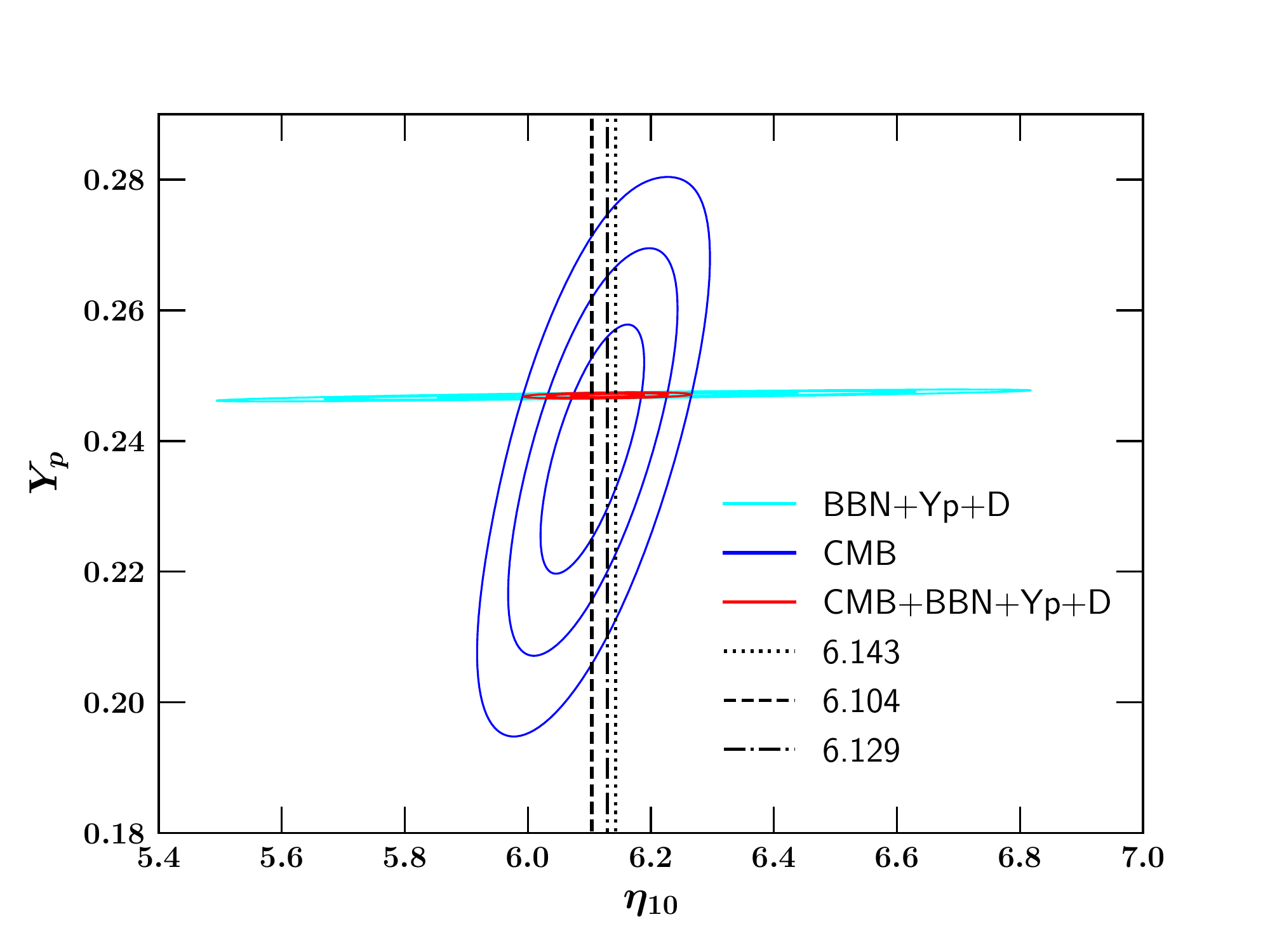}
        \hskip -.2in
    \includegraphics[width=0.51\textwidth]{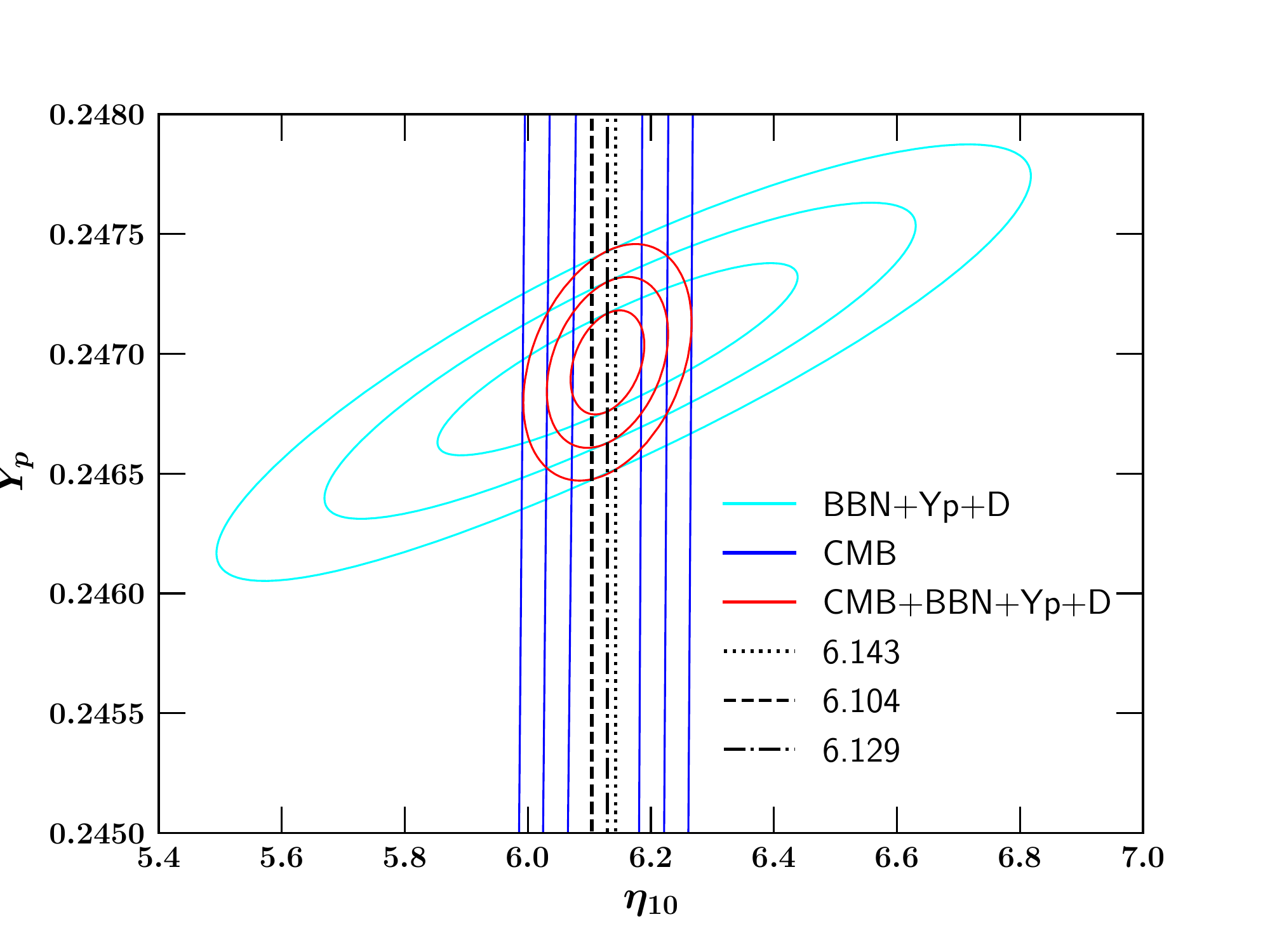}
    \caption{Likelihood functions using the FOYY \dpg rate  projected onto the $(\eta_{10},Y_p)$ plane. The right panel focuses on a more narrow range of values of $Y_p$.  }
    \label{fig:Deta}
\end{figure}

This behavior (or lack thereof) can be understood by the likelihood functions shown in Fig.~\ref{fig:Deta}.
Here we show projections of the likelihood functions onto the $(Y_p$, $\eta_{10})$ plane. The tilted blue ellipses in the left panel show the 68.3\%, 95.4\%, and 99.7\% confidence level contours for the the CMB-alone likelihood function, ${\mathcal L}_{\rm CMB}(\eta,Y_p)$. This is centered on $\eta_{10} = 6.104$. 
Ignoring the CMB, we show the convolved BBN+Obs likelihood by the set of cyan ellipses.
These are more easily seen in the zoomed-in version shown in the right panel. Despite the precision in the observed D/H abundance, the mean value for $\eta_{10}$ found from this likelihood function is $6.143 \pm 0.190$ \cite{foyy}. Thus the observed abundance, which for the FOYY \dpg rate is lower than the calculated abundance, tries to drive the value of $\eta$ up, 
when combined with the CMB. 
However the large uncertainty in $\eta$ in the BBN+Obs likelihood means that the pull is very weak.  Indeed, for CMB+BBN+Obs the mean value
in $\eta_{10}$ remains essentially the same at 6.129, shifting by $<0.001$.

\begin{figure}[!htb]
    \centering
    \includegraphics[width=0.51\textwidth]{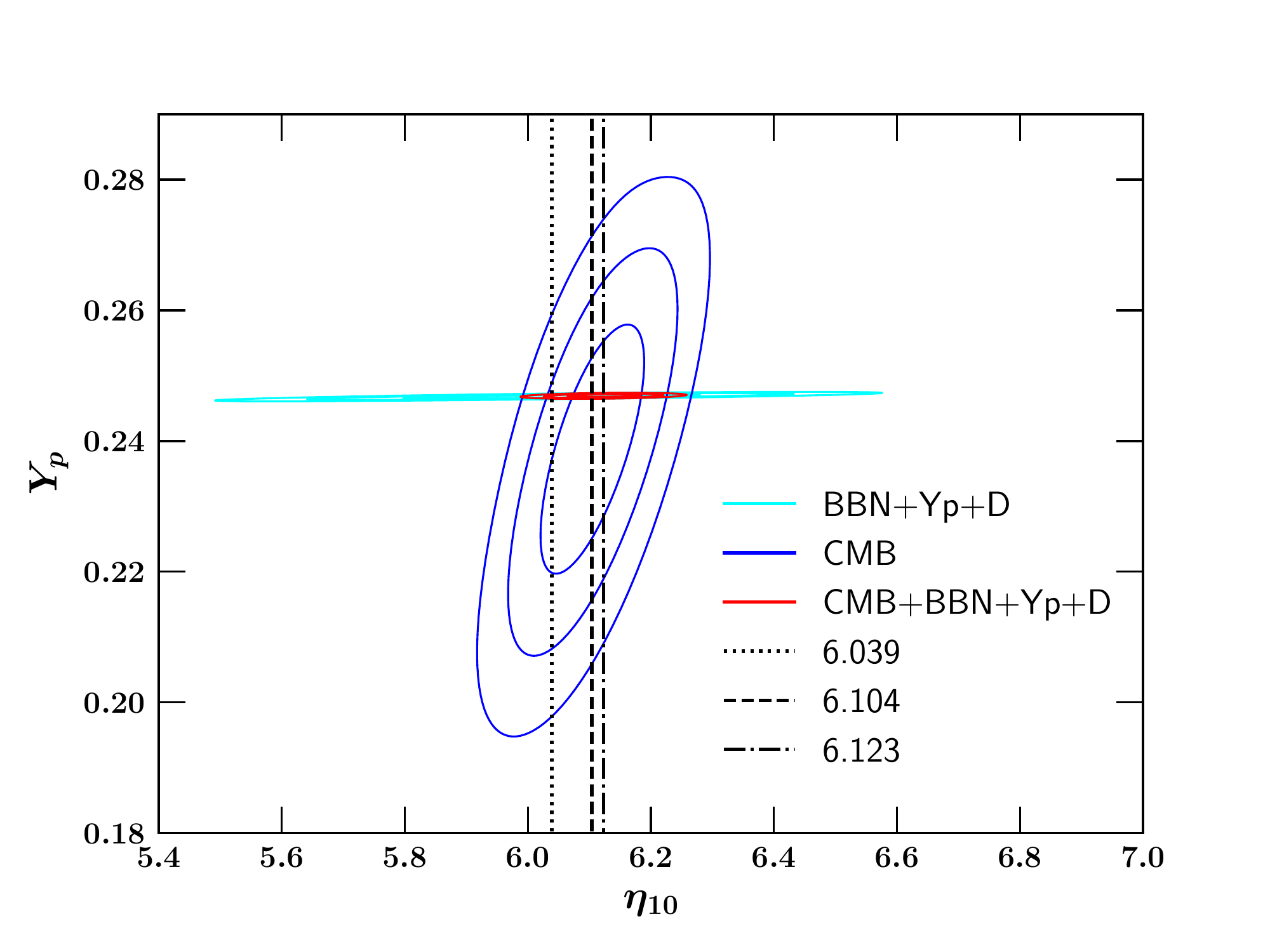}
    \hskip -.2in
    \includegraphics[width=0.51\textwidth]{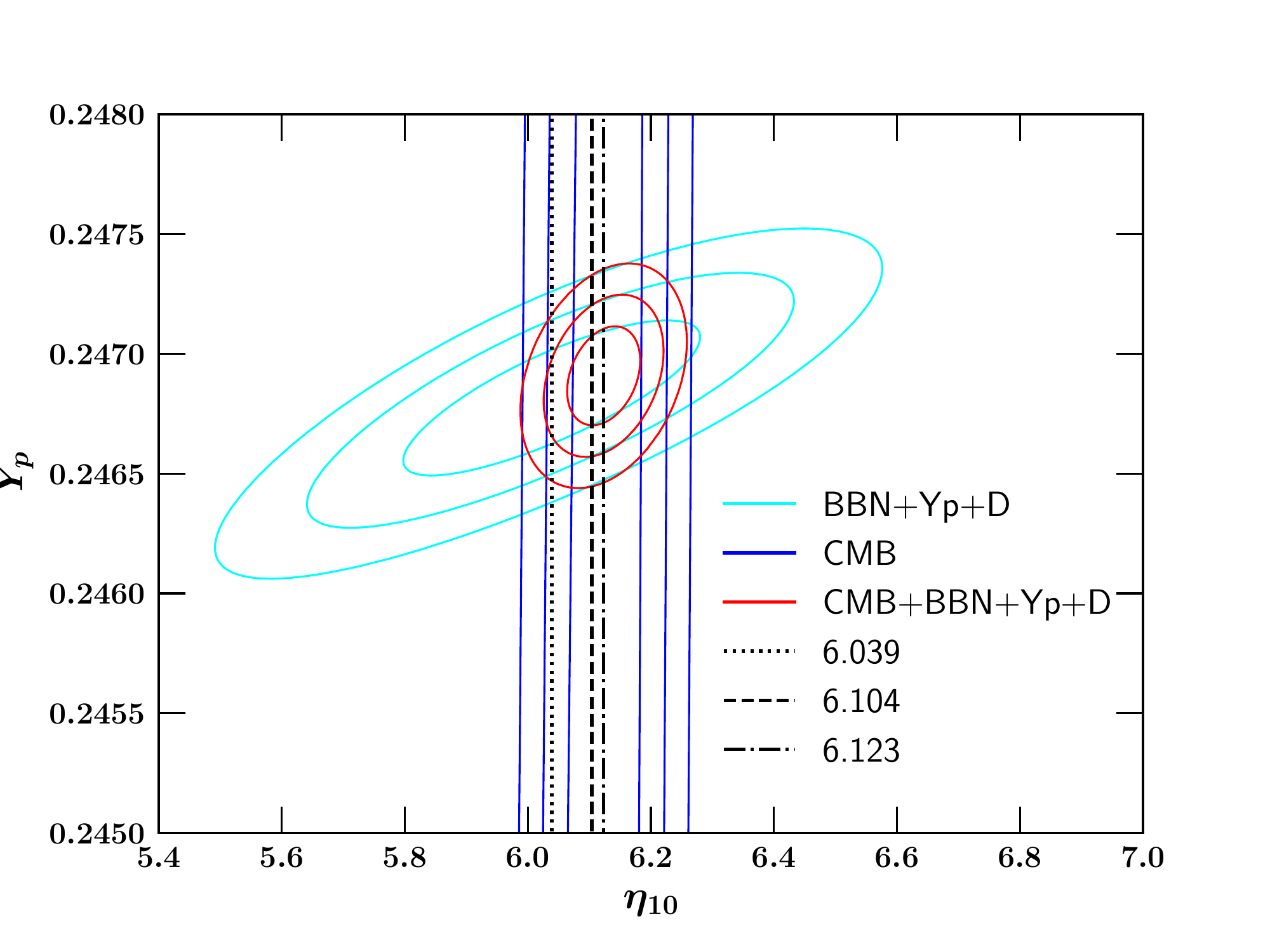}
    \caption{Likelihood functions using our new averaged \dpg rate  projected onto the $(\eta_{10},Y_p)$ plane. The right panel focuses on a more narrow range of values of $Y_p$.  }
    \label{fig:Detanew}
\end{figure}

We show in Fig.~\ref{fig:Detanew}, a similar projection of the likelihood functions when using the newly averaged \dpg rate which includes the LUNA measurement \cite{mossa2}. As one can see, the BBN+Obs ellipses remain elongated due to the residual uncertainty in the rates which 
determine D/H. In addition we see that they are 
shifted to lower values of $\eta$. Indeed, these are are now centered around $\eta_{10} = 6.04$ below both the CMB and BBN+CMB likelihoods. Nevertheless, 
the because of the relatively large uncertainty in $\eta$, the combined likelihood, while shifted to lower values of $\eta_{10}$, shifts only to 6.123 (from 6.129).

\begin{figure}[!htb]
    \centering
    \includegraphics[width=0.75\textwidth]{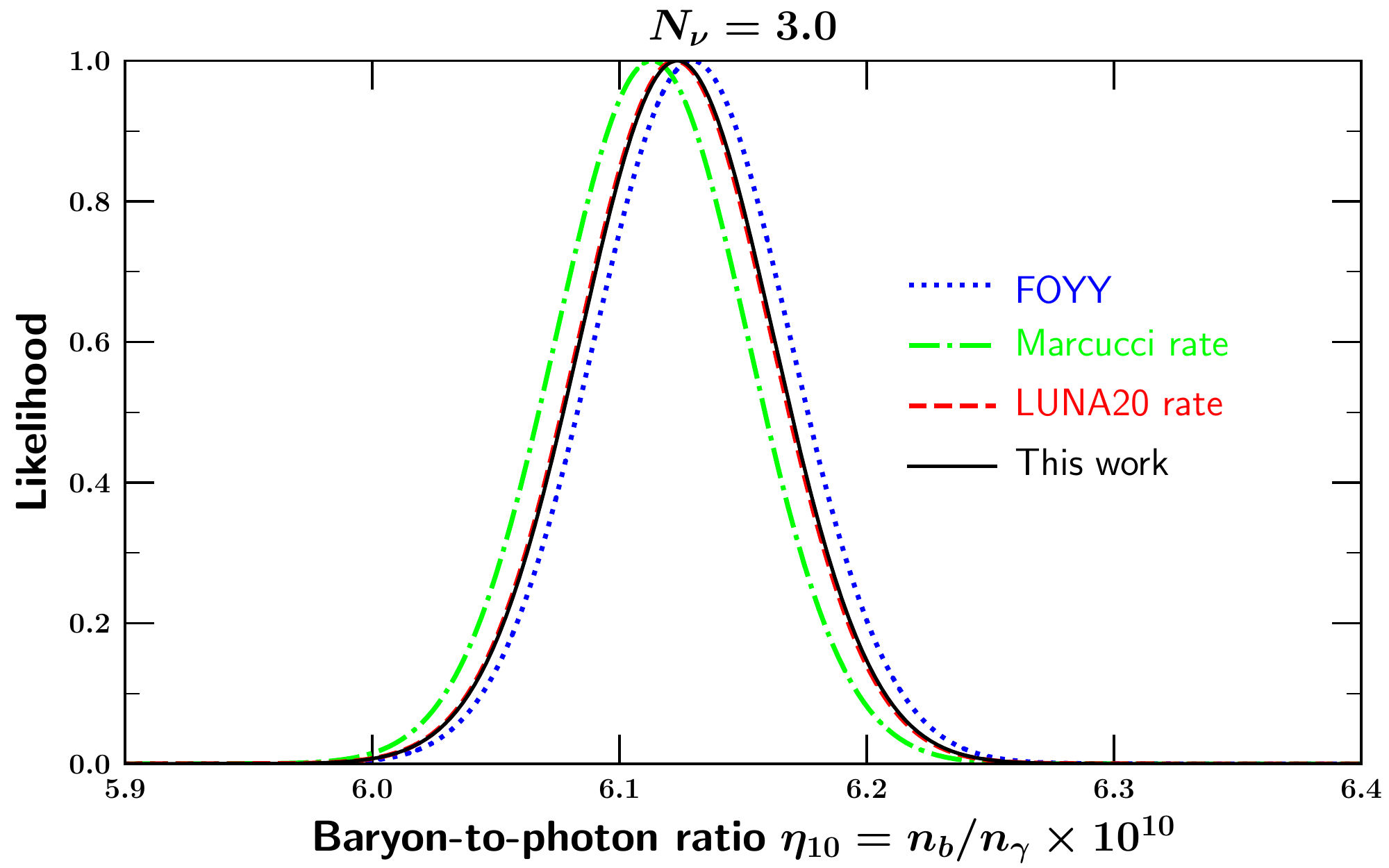}
    \caption{Baryon-to-photon ratio determinations using the likelihood function defined in Eq.~(\ref{CMB-OBS-eta}) for the four choices of the \dpg rate considered. }
    \label{fig:SBBN-eta-baseline}
\end{figure}

Our total combined likelihood distribution marginalized to give a function of $\eta$
can be expressed as
\beq
{\mathcal L}_{\rm BBN+CMB+Obs}(\eta) \propto \int 
{\mathcal L}_{\rm CMB}(\eta,Y_p)
  {\mathcal L}_{\rm BBN}(\eta;X_i) \
  {\mathcal L}_{\rm Obs}(X_i) \ \prod_i dX_i \, ,
  \label{CMB-OBS-eta}
\eeq
which is plotted in Fig.~\ref{fig:SBBN-eta-baseline}. Shown are the results for our four choices of the \dpg rate. 
Its mean and standard deviation are given in Table \ref{tab:eta}.
Our final combined value for the baryon-to-photon ratio, is therefore 
\beq
\eta = (6.123 \pm 0.039) \times 10^{-10} \qquad \omb = 0.02237 \pm 0.00014 \, .
\eeq
Also given in the table is the value of $\eta_{10}$ at the peak of the likelihood distribution which in these cases equals the mean.

\begin{table}[!htb]
\caption{Constraints on the baryon-to-photon ratio, using four choices for the \dpg rate.  We have marginalized over $Y_p$ to create 1D $\eta$ likelihood distributions. Given are both the mean (and its uncertainty)
as well as the value of $\eta$ at the peak of the distribution.
\label{tab:eta}
}
\vskip .2in
\begin{tabular}{|l|c|c|}
\hline
 \dpg rate & mean $\e10$ & peak $\e10$ \\
\hline
FOYY~\cite{foyy} & $6.129\pm 0.040$ & 6.129 \\
\hline
Theory~\cite{marc} & $6.113\pm 0.039$ & 6.113 \\
\hline
LUNA20~\cite{mossa2} & $6.123\pm 0.039$ & 6.123 \\
\hline
\hline
This Work & $6.123\pm 0.039$ & 6.123 \\
\hline
\end{tabular}
\end{table}

\section{Results for variable $N_\nu$}
\label{sec:Nnot3}

\begin{center}
\begin{figure}[!htb]
\includegraphics[width=0.80\textwidth]{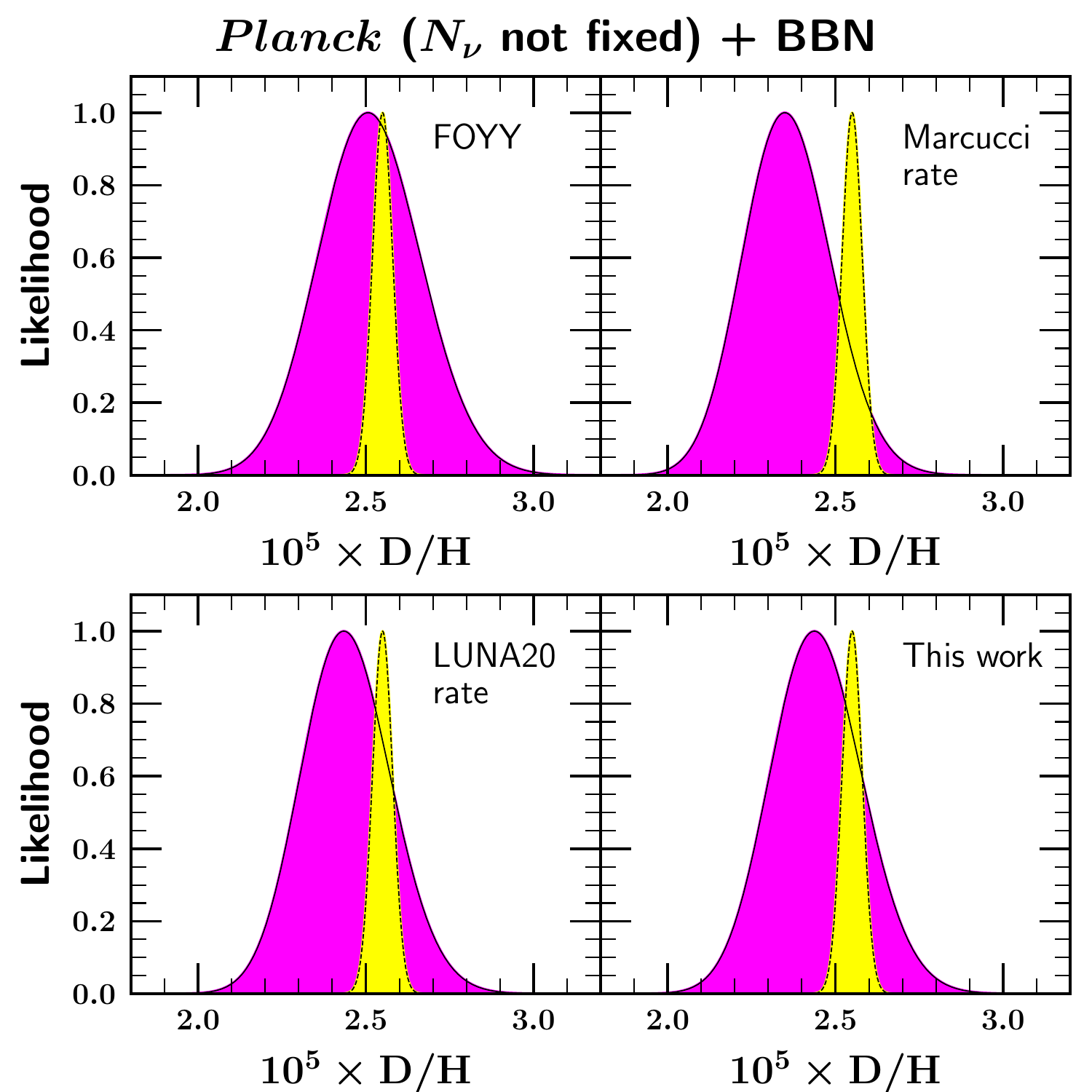}
\caption{ As in Fig. \ref{fig:2x2abs_2d}, the
predictions for D/H using the CMB determination of the 
cosmic baryon density when $N_\nu$ is not fixed.   
}
\label{fig:2x2abs_3d}
\end{figure}
\end{center}

We turn now to the effect of the new \dpg rate when we allow $N_\nu$ to differ from its Standard Model value of 3. We can repeat the above analysis substituting ${\mathcal L}_{\rm NCMB}(\eta,Y_p,N_\nu)$ for ${\mathcal L}_{\rm CMB}(\eta,Y_p)$ and ${\mathcal L}_{\rm NBBN}(\eta,N_\nu;X_i)$ for ${\mathcal L}_{\rm BBN}(\eta;X_i)$. Integrating these generalized likelihood functions over $\eta$ as in Eq.~(\ref{CMB+BBN}), we obtain the likelihood distributions for D/H shown in Fig.~\ref{fig:2x2abs_3d} for our four choices of \dpg rates. 
As we would expect, the higher \dpg rates (relative to the one used in FOYY) causes
the BBN likelihood function for D/H (shown in purple) to shift to the left (towards smaller values of D/H). In fact, when $N_\nu$ is not fixed, the shift to lower D/H is greater than we saw in Fig.~\ref{fig:2x2abs_2d}, for reasons which will become clear below.

The BBN+CMB likelihoods in Fig.~\ref{fig:2x2abs_3d}
are summarized by
the predicted abundances for D/H given in Table~\ref{tab:DH3d}
where again the central values give the mean,
and the error gives the $1\sigma$ variance and
the final column gives the value at the peak of the distribution shown in the purple shaded distributions in Fig.~\ref{fig:2x2abs_3d}. For completeness, we also provide in Table~\ref{tab:YLi3d},
the \he4 and \li7 abundances for each case. 

\begin{table}[!htb]
\caption{The mean and peak values of D/H for each of the 
adopted rates for \dpg when $N_\nu$ is not held fixed.
\label{tab:DH3d}
}
\begin{tabular}{|l|c|c|}
\hline
 \dpg rate & mean D/H $\times 10^5$ & peak D/H $\times 10^5$ \\
\hline
FOYY~\cite{foyy} & $2.514\pm0.152$ & $2.506 $ \\
\hline
Theory~\cite{marc} & $2.359\pm 0.131$ & $2.349$ \\
\hline
LUNA20~\cite{mossa2} & $2.444\pm 0.133$ & $2.434$ \\
\hline
\hline
This Work & $2.447 \pm 0.137$ & $2.437$ \\
\hline
\end{tabular}
\end{table}

\begin{table}[!htb]
\caption{The mean and peak values of $Y_p$ and Li/H for each of the 
adopted rates for \dpg when $N_\nu$ is not held fixed.
\label{tab:YLi3d}
}
\begin{tabular}{|l|c|c|}
\hline
 \dpg rate & mean $Y_p$ & peak $Y_p$ \\
\hline
FOYY~\cite{foyy}  & $0.24408\pm 0.00407$ & $0.24440 $ \\
\hline
Theory~\cite{marc} & $0.24409\pm0.00408$ & $0.24441$ \\
\hline
LUNA20~\cite{mossa2} & $0.24408\pm0.00408$ & $0.24441$ \\
\hline
\hline
This Work & $0.24408\pm0.00408$ & $0.24441$ \\
\hline
\hline
 \dpg rate & mean Li/H $\times 10^{10}$ & peak Li/H $\times 10^{10}$ \\
\hline
FOYY~\cite{foyy}  & $4.78\pm 0.74$ & $4.78$ \\
\hline
Theory~\cite{marc} & $5.34 \pm 0.77$ & $5.33$ \\
\hline
LUNA20~\cite{mossa2} & $5.02 \pm 0.72$ & $5.00$ \\
\hline
\hline
This Work & $5.01\pm 0.73$ & $5.00$ \\
\hline
\end{tabular}
\end{table}

\begin{figure}[!htb]
    \centering
    \includegraphics[width=0.95\textwidth]{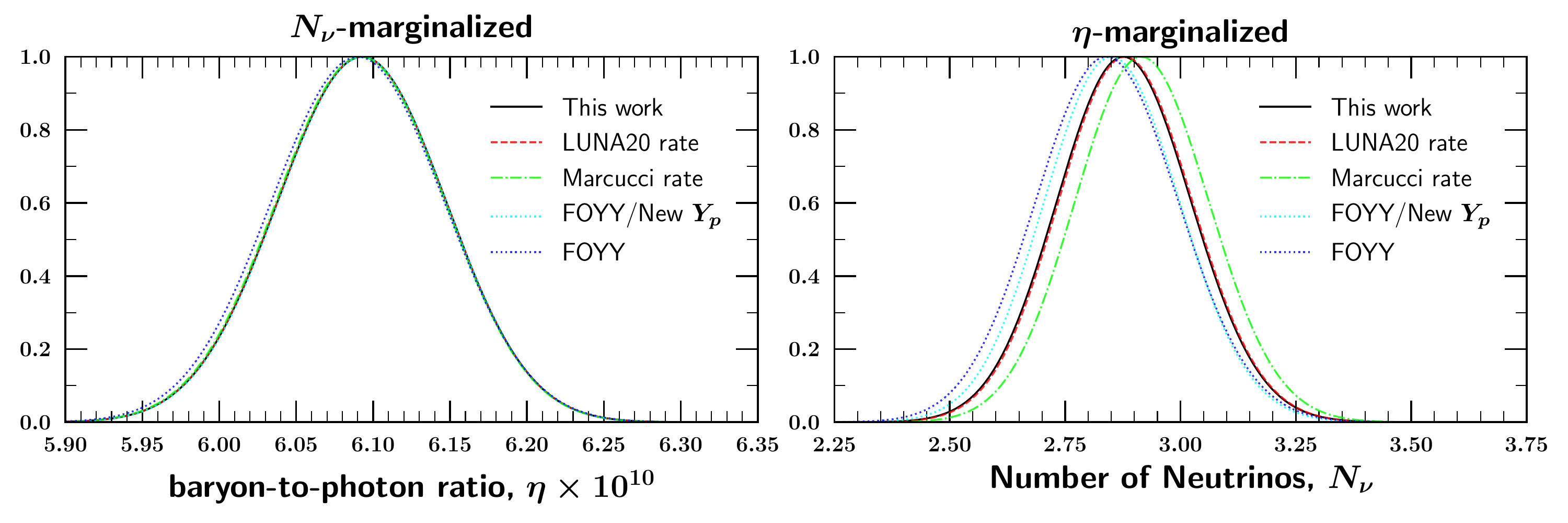}
    \caption{(Left) As in Fig.~\ref{fig:SBBN-eta-baseline}, but where $N_\nu$ is allowed to vary, and results marginalized over it.  (Right) The likelihood as a function of $N_\nu$, which is marginalized over $\eta$. }
    \label{fig:NBBN-eta-baseline}
\end{figure}

In this case, we can obtain 1-dimensional likelihood functions by integrating with respect to $N_\nu$ or $\eta$ given by
\beq
{\mathcal L}_{\rm NBBN+NCMB+Obs}(\eta) \propto \int 
{\mathcal L}_{\rm NCMB}(\eta,Y_p, N_\nu)
  {\mathcal L}_{\rm NBBN}(\eta,N_\nu;X_i) \
  {\mathcal L}_{\rm Obs}(X_i) \ \prod_i dX_i dN_\nu \, ,
  \label{NCMB-NBBN-NOBS-eta}
\eeq
or 
\beq
{\mathcal L}_{\rm NBBN+NCMB+Obs}(N_\nu) \propto \int 
{\mathcal L}_{\rm NCMB}(\eta,N_\nu,Y_p)
  {\mathcal L}_{\rm NBBN}(\eta,N_\nu;X_i) \
  {\mathcal L}_{\rm Obs}(X_i) \ \prod_i dX_i d\eta \, .
  \label{CMB-BBN-OBS-nnu}
\eeq
These are shown in the left and right panels of Fig.~\ref{fig:NBBN-eta-baseline} respectively. Almost independent of the \dpg rate, we see that the best fit solution for $\eta_{10}$ is lower
and we find $\eta_{10} = 6.092 \pm 0.054$. 
The mean, width, and peak of each of these likelihood functions are summarized in Table~\ref{tab:etannu}. 
Also included in this table is the result based the FOYY rate, but with an updated value for $Y_p$ from \cite{abopss}. 
While the value of $\eta$ is only marginally increased by the updated $Y_p$, the mean value of $N_\nu$ is increased and its uncertainty is decreased.  The new value of $Y_p$ is
used in all subsequent results shown. 
This shift downwards in 
$\eta$ is predominantly due to the CMB likelihood as discussed in FOYY. Similarly, the most likely values for $N_\nu$
are also slightly below 3.0.
Using the new rate for \dpg, when BBN and CMB and observations are combined
we find 
\beq
N_\nu = 2.880 \pm 0.144
\eeq
giving 68\% and 95\% confidence level upper limits to $N_\nu -3$ of 0.024 and 0.168 respectively.

\begin{table}[!htb]
\caption{The marginalized most-likely values and central 68.3\% confidence limits on the baryon-to-photon ratio and effective number of neutrinos, for each choice of the \dpg rate. 
\label{tab:etannu}
}
\vskip.1in
\begin{tabular}{|l|c|c|c|c|}
\hline
 \dpg rate & mean $\eta_{10}$ & peak $\eta_{10}$ & mean $N_\nu$ & peak $N_\nu$ \\
\hline
FOYY~\cite{foyy}   & $6.090 \pm 0.055$ & 6.090 & $2.843 \pm 0.154$ & 2.839 \\
\hline
updated $Y_P$~\cite{foyy,abopss}  & $6.093 \pm 0.054$ & 6.093 & $2.855 \pm 0.146$ & 2.851 \\
\hline
Theory~\cite{marc} & $6.092\pm 0.054$ & 6.092 & $2.918 \pm 0.144$ & 2.915 \\
\hline
LUNA20~\cite{mossa2} & $6.092 \pm 0.054$ & 6.093 &  $2.883 \pm 0.144$ & 2.879 \\
\hline
\hline
This Work & $6.092 \pm 0.054$ & 6.093 & $2.880 \pm 0.144$ & 2.876 \\
\hline
\end{tabular}
\end{table}

These results allow us to understand the downward shift in D/H seen in Fig.~\ref{fig:2x2abs_3d}.
While, the lower value of $\eta$ (6.09 relative to 6.12) produces a 0.8\% increase in D/H,
this is overcompensated by a 1.6\%
shift down in D/H due to the lower value in $N_\nu$, since D/H scales as $N_\nu^{0.405}$ as given in the appendix. 

\section{Summary}
\label{sec:summary}

Predictions from BBN are limited by the precision of the nuclear rates used in BBN calculations. As such, the new rates for \dpg provided by the LUNA collaboration are indeed welcome. In addition to providing a higher degree of accuracy, they also help resolve the discrepancy between older experimental rates, and theoretical calculations, suggesting that more work on the theory is needed to explain the new rates.

The new rates reaffirm the excellent agreement 
between BBN/CMB calculated light element abundances, and their observational determinations. At the fixed value of $\eta$ as determined in this work and shown in the last line of Table~\ref{tab:eta}, we compare our results with previous results.

\begin{table}[ht]
\caption{Comparison of BBN Results
\label{tab:re}
}
\begin{tabular}{|l|c|c|c|c|c|c|c|}
\hline
 & $\eta_{10}$ & $N_{\nu}$ & $Y_p$ & D/H & \he3/H & \li7/H  \\
\hline
Ref. \cite{CFOY}& 6.10 & 3 & 0.2470 & 2.579 $\times$ $10^{-5}$& 0.9996 $\times$ $10^{-5}$ &  $4.648 \times$ $10^{-10}$  \\
\hline
Ref. \cite{coc18}& 6.091 & 3 & 0.2471 & 2.459 $\times$ $10^{-5}$& 1.074 $\times$ $10^{-5}$ &  $5.624 \times$ $10^{-10}$  \\
\hline
Ref. \cite{foyy} & 6.129 & 3 & 0.2470 & 2.559 $\times$ $10^{-5}$& 0.9965 $\times$ $10^{-5}$ &  $4.702 \times$ $10^{-10}$  \\
\hline
This Work & 6.123 & 3 & 0.2470 & 2.493 $\times$ $10^{-5}$& 1.033 $\times$ $10^{-5}$ &  $4.926 \times$ $10^{-10}$  \\
\hline
\end{tabular}
\end{table}

Table \ref{tab:re} compares our results with those of earlier work.
The values quoted are not from a full Monte Carlo analysis
but rather results for a single value of $\eta$, and the reaction
rates at their central value.  
The abundance differences are the result of different nuclear rates.  
Our new D/H abundance is lower than our earlier results in
FOYY \cite{foyy} and in ref.~\cite{CFOY}
due to the increased destruction via \dpg.  While it appears that our new
result for D/H is closer to that of the Paris group \cite{coc18},
these are evaluated at different values of $\eta$. At the same
value of $\eta$ our D/H remains significantly higher, as the \dpg rate employed in
\cite{coc18} lies above the LUNA data.
Finally, we see that for the same $\eta$ our $Y_p$
and \he3 abundances are quite similar due to their weak dependence on 
this rate.

\begin{center}
\begin{figure}[!htb]
\includegraphics[width=0.7\textwidth]{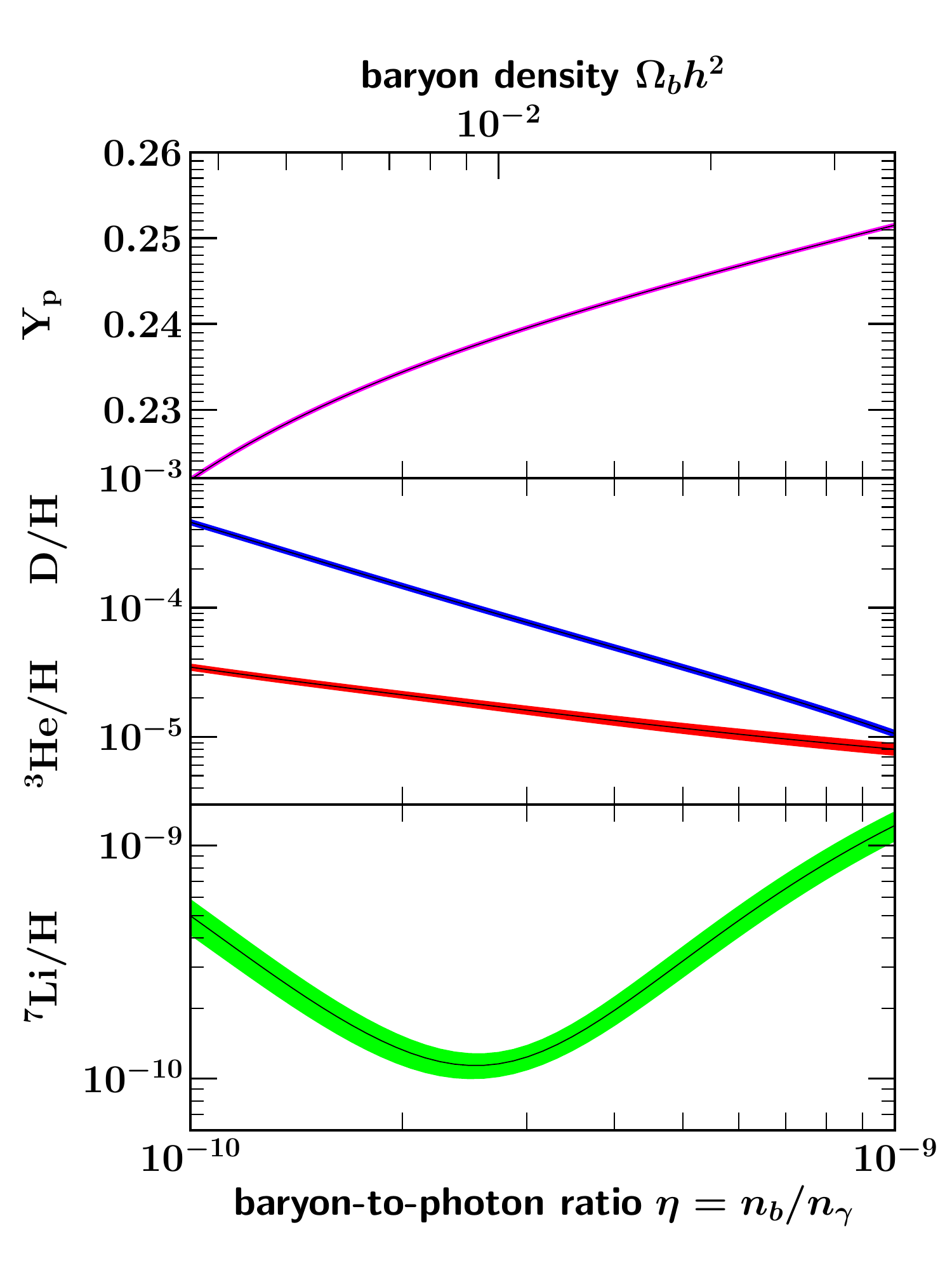}
\caption{
Primordial abundances of the light nuclides as a function of
cosmic baryon content, as predicted by SBBN (``Schramm plot'').
These
results assume $N_\nu = 3$ and the current measurement of the neutron lifetime $\tau_n = 879.4 \pm 0.6$ s. 
Curve widths show $1\sigma$ errors.
\label{fig:schramm}
}
\end{figure}
\end{center}

\begin{figure}[!htb]
    \centering
    \includegraphics[width=0.7\textwidth]{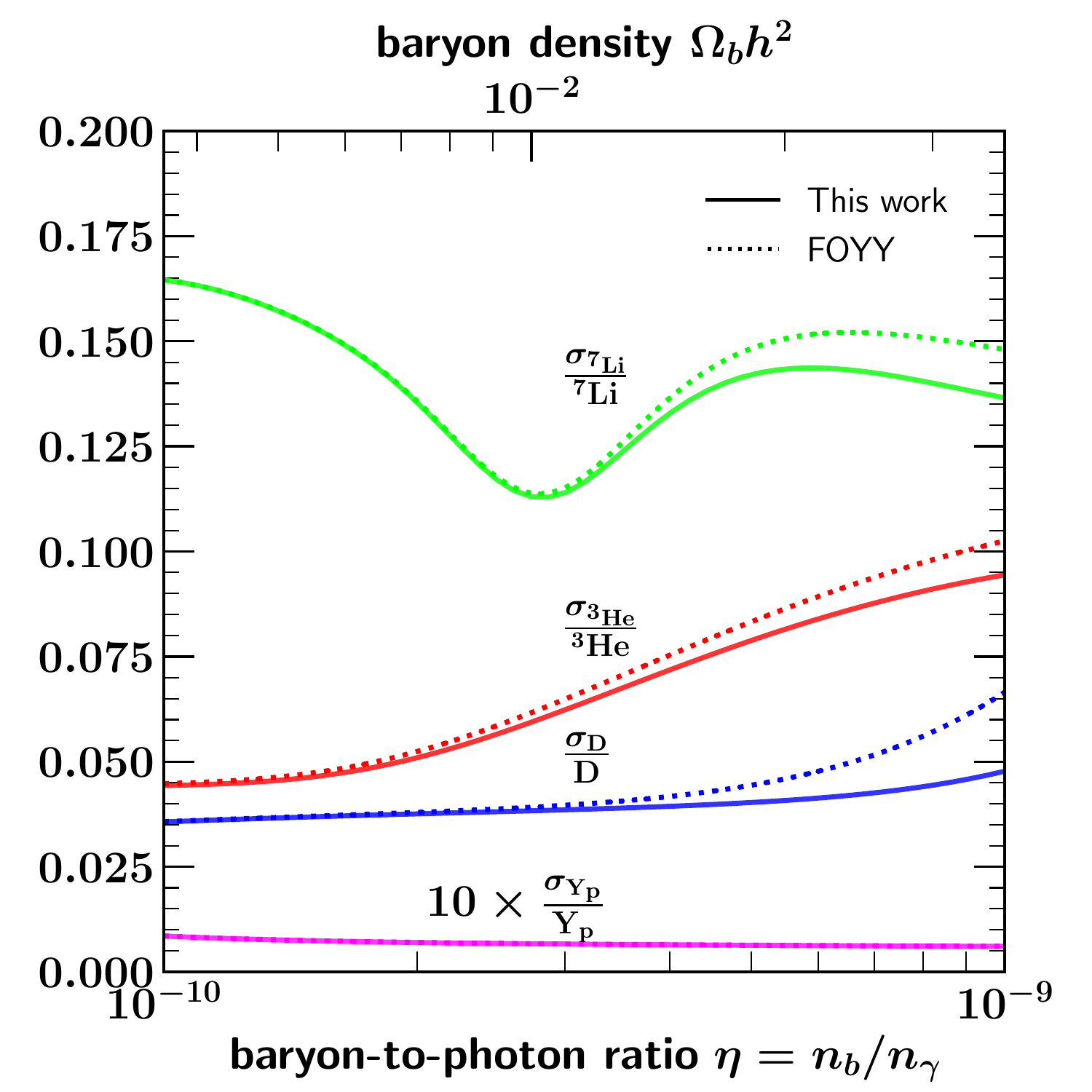}
    \caption{Fractional uncertainties in the light element abundance predictions shown in Fig.~\ref{fig:schramm}.  For each species $i$, we plot ratio of the standard deviation $\sigma_i$ to the mean $\mu_i$, as a function of baryon-to-photon ratio. The relative uncertainty of the \he4 abundance has been multiplied by a factor of 10. }
    \label{fig:schrammerrors}
\end{figure}

Fig.~\ref{fig:schramm} summarizes our results with a Schramm plot on our newly averaged \dpg rate.
The thickness of each curve depicts the $\pm 1 \sigma$ spread in the predicted abundances.  The relative uncertainty,
(the thickness of the curves, relative to the central value) is shown more clearly in Figure~\ref{fig:schrammerrors}. As one can see, the \li7 abundance remains the most uncertain, with a relative uncertainty of approximately 13\%. 
The uncertainty in the deuterium abundance ranges from 3.5-5\%,
while the calculated uncertainty in \he4 is only 0.1\% (note that it is amplified by a factor of 10 in the figure for clarity). 
Comparison with the FOYY results shown by the dotted curves confirms the substantial improvement due to the LUNA measurements.

As we have seen, the uncertainty in the BBN prediction of D/H has decreased, yet still 
falls short of the precision of the observations.
It is thus important to re-evaluate the error budget for primordial deuterium.  
As seen in the Appendix,
in addition to \dpg, deuterium has strong sensitivity to
the $d(d,n)\he3$ and $d(d,p)t$ rates.
To quantify this effect, we have run a Monte Carlo calculation
at fixed $\eta_{10}=6.123$ to evaluate the contribution $\sigma_i({D/H})$
to the D/H uncertainty from individual reactions $i$
with all other rate uncertainties set to zero;
these are shown in the second column of Table \ref{tab:errbudget}.
In the third column of Table \ref{tab:errbudget},
we show the D/H uncertainty when {\em only}
the contribution from reaction $i$ is set to zero,
denoted $\sigma_{{\rm omit} \; i}({\rm D/H})$.
One can verify that the total uncertainty very closely follows
the simple error propagation result
for the sum of uncorrelated random variables:
$\sigma_{\rm tot}^2 = \sum_i \sigma_i^2$.
Similarly, $\sigma_{\rm tot}^2 = \sigma_i^2 + \sigma_{{\rm omit} \; i}^2$ to an excellent approximation.

Table \ref{tab:errbudget} summarizes the new
landscape of D/H nuclear uncertainties and 
points the way to future progress.
We see that \dpg\ is now well under control: 
it barely contributes to the total D/H error.  Indeed, 
even letting the uncertainty go to zero would barely improve the final D/H precision.
This is a testament to the power of the LUNA result.
Thus, further improvement in D/H turns to 
other reactions.  In particular, $d(d,n)\he3$
now clearly dominates the error budget,
with $d(d,p)t$ in somewhat distant second place.
Fortunately, both reactions have the same initial
state, and so both can be accessible in the same
experiment.  From the table we see that a factor of $\sim 2$ improvement in uncertainty
would bring the $d(d,n)$ contribution to $\sigma({\rm D/H})$ down the level of the new \dpg\ results.

\begin{table}[]
    \centering
    \caption{Deuterium Uncertainty Contributions at $\eta_{10}=6.123$}
    \begin{tabular}{|c|c|c|}
    \hline 
    Reaction $i$ &  $10^5 \ \sigma_i({\rm D/H})$  &
    $10^5 \ \sigma_{{\rm omit} \; i}({\rm D/H})$ \\
    \hline
    \dpg  & 0.036 & 0.097 \\
    \hline
    $d(d,n)\he3$  & 0.081 & 0.065 \\
    \hline
    $d(d,p)t$ & 0.054 & 0.089 \\
    \hline 
    $\he3(d,p)\he4$ & 0.002 & 0.103 \\
    \hline
    $p(n,\gamma)d$ & 0.002 & 0.103 \\
    \hline
    $\he3(n,p)t$ & 0.002 & 0.103 \\  
    \hline \hline
    all &  0.103 & -- \\
    \hline  
    \end{tabular}
    \label{tab:errbudget}
\end{table}

We have thus seen that the new LUNA measurements have substantially improved the experimental underpinnings of primordial deuterium production.
The \dpg\ rate is now well-determined in the BBN energy range,
and contributes small part of the D/H error budget. 
The D/H theory prediction is now significantly sharper, though it has
still has not caught up to observational precision.
These improvements are important for both standard and non-standard BBN.
For standard BBN, the constraints on the
baryon density tighten and thus afford a sharper comparison with CMB.
For NBBN, we find notably stronger bounds on $N_\nu$.

Future work can capitalize on these improvements and push for further
precision.
Because the BBN+CMB prediction errors still lag the observations,
we urge new, precise measurements of the $d(d,n)$ and $d(d,p)$ reactions
in the BBN energy range.
The planned next generation CMB-Stage IV experiment 
\cite{CMB-S4-RefDes}
should substantially improve the CMB determination of $N_\nu$ and $Y_p$, increasing the power of the comparison with BBN
and possibly detecting the $N_{\rm eff}-N_\nu=0.045$ difference, thus probing effects of neutrino heating during $e^\pm$ annihilation.  Finally, of course the
lithium problem continues to awaits a firm solution.

{\it Note added:}
As this paper was in the final stages of preparation, we became aware
of two works similar to our own.  
Ref.~\cite{Pisanti2020} builds on their work with
LUNA \cite{mossa2}, coming to similar conclusions as ours for SBBN and NBBN.
In Ref.~\cite{Pitrou2020}, the Paris group performs their own analysis of the impact of the LUNA measurements on Standard BBN with $N_\nu=3$. Their nuclear rates
make use of theory, including for \dpg, giving a deuterium abundance $({\rm D/H})_{\rm BBN+CMB} = (2.439 \pm 0.037) \times 10^{-5}$. This is substantially lower than both our result
and the observations, and consequences of this mismatch are discussed.
That these conclusions are so different emphasizes the continuing central role of nuclear reactions and their analysis for BBN.

\section*{Appendix}
\label{sect:app}

In this appendix, we update the fractional dependence of the light element abundances to
$\eta$, $N_\nu$, the gravitational constant,
the neutron mean life, and a selection of nuclear rates. For each element we give
\beq
\label{eqn:sens}
X_i = X_{i,0}\prod_n \left(\frac{p_n}{p_{n,0}}\right)^{\alpha_n} \, ,
\eeq
where $X_i$ represents either the helium mass fraction or the abundances of the other
light elements by number. The $p_n$ represent input quantities to the BBN calculations, whose power-law scaling is the log derivative
$\alpha_i = \partial \ln X_i/\partial \ln p_i$. 
Below are the scalings with these inputs, as defined in detail in FOYY and ref.~\cite{CFOY}. 
\beqar
Y_p &=& 0.24695\!\left(\frac{\e10}{6.123}\right)^{0.039}\!\!\left(\frac{N_\nu}{3.0}\right)^{0.163}\!\!\left(\frac{G_N}{G_{N,0}}\right)^{0.357}\!\!\left(\frac{\tau_n}{879.4 s}\right)^{0.729} \nonumber \\
 &\times& \left[ p(n,\gamma)d\right]^{0.005}\left[ d(d,n)\he3\right]^{0.006}\left[ d(d,p)t\right]^{0.005} 
\label{yfit}
\eeqar
\beqar
\frac{\rm D}{\rm H} &=& 2.493\!\times\! 10^{-5}\!\left(\frac{\e10}{6.123}\right)^{-1.634}\!\!\left(\frac{N_\nu}{3.0}\right)^{0.405}\!\!\left(\frac{G_N}{G_{N,0}}\right)^{0.974}\!\!\left(\frac{\tau_n}{879.4 s}\right)^{0.418} \nonumber  \\
&\times& \left[ p(n,\gamma)d\right]^{-0.197}\left[ d(d,n)\he3\right]^{-0.529}\left[ d(d,p)t\right]^{-0.468} \nonumber \\
&\times& \left[d(p,\gamma)\he3\right]^{-0.344}\left[\he3(n,p)t\right]^{0.024}\left[\he3(d,p)\he4\right]^{-0.014} 
\eeqar
\beqar
\frac{\he3}{\rm H} &=& 1.033\!\times\! 10^{-5}\!\left(\frac{\e10}{6.123}\right)^{-0.564}\!\!\left(\frac{N_\nu}{3.0}\right)^{0.135}\!\!\left(\frac{G_N}{G_{N,0}}\right)^{0.324}\!\!\left(\frac{\tau_n}{879.4 s}\right)^{0.139} \nonumber \\
&\times& \left[ p(n,\gamma)d\right]^{0.087}\left[ d(d,n)\he3\right]^{0.198}\left[ d(d,p)t\right]^{-0.257} \nonumber  \\
&\times& \left[d(p,\gamma)\he3\right]^{0.395}\left[\he3(n,p)t\right]^{-0.166}\left[\he3(d,p)\he4\right]^{-0.759}\left[t(d,n)\he4\right]^{-0.008} 
\eeqar
\beqar
\frac{\li7}{\rm H} &=& 4.926\!\times\! 10^{-10}\!\left(\frac{\e10}{6.123}\right)^{2.118}\!\!\left(\frac{N_\nu}{3.0}\right)^{-0.286}\!\!\left(\frac{G_N}{G_{N,0}}\right)^{-0.732}\!\!\left(\frac{\tau_n}{879.4 s}\right)^{0.429} \nonumber  \\
&\times& \left[ p(n,\gamma)d\right]^{1.308}\left[ d(d,n)\he3\right]^{0.677}\left[ d(d,p)t\right]^{0.062} \nonumber  \\
&\times& \left[d(p,\gamma)\he3\right]^{0.623}\left[\he3(n,p)t\right]^{-0.266}\left[\he3(d,p)\he4\right]^{-0.750}\left[t(d,n)\he4\right]^{-0.021} \nonumber  \\
&\times& \left[\he3(\alpha,\gamma)\be7\right]^{0.965}\left[\be7(n,p)\li7\right]^{-0.692}\left[\li7(p,\alpha)\he4\right]^{-0.052}\left[t(\alpha,\gamma)\li7\right]^{0.027} \nonumber  \\
&\times&\left[\be7(n,\alpha)\he4\right]^{-0.001}\left[\be7(d,p)\he4\he4\right]^{-0.008}
\label{li7fit}
\eeqar

\section*{Acknowledgments}
The work of K.A.O.~was supported in part by DOE grant DE-SC0011842  at the University of
Minnesota.

\end{document}